\documentclass{article}

\usepackage{amsthm}
\usepackage{amsmath}
\theoremstyle{plain}
\newtheorem{lemma}{Lemma}

\usepackage{afterpage}
\usepackage{microtype}
\usepackage{graphicx}
\usepackage{subfigure}
\usepackage{booktabs} 
\usepackage{graphicx}
\usepackage{tikz}
\usepackage{enumitem}
\usepackage{xparse}
\usepackage{pifont}
\usepackage{makecell}
\usepackage{hyperref}
\usepackage{xurl}
\usepackage[normalem]{ulem}

\usepackage{cleveref}
\newif\ifdraft
\draftfalse

\newif\ifdiff
\difftrue

\newcommand{\sys}{FCP}
\newcommand{\sota}{state-of-the-art}

\newcommand{\ion}[1]{\ifdraft{\noindent{\textcolor{blue}{\bf \fbox{IS} {\it#1}}}}\fi}

\newcommand{\CD}{Block Distributor}
\newcommand{\cd}{block distributor}
\newcommand{\cs}{communication planner}
\newcommand{\CS}{Communication Planner}
\newcommand{\tr}{transparent reshuffler}
\newcommand{\TR}{Transparent Reshuffler}
\newcommand{\ptop}{peer-to-peer}

\newcommand{\fig}[1]{Figure ~\ref{#1}}

\newcommand{\refsec}[1]{\S~\ref{#1}}
\newcommand{\MyPara}[1]{\vspace{.2em}\noindent\textbf{#1}~}

\NewDocumentEnvironment{draftfig}{O{htbp}}{%
  \begin{figure}[#1]
    \centering
    \begingroup
    \let\oldincludegraphics\includegraphics
    \renewcommand{\includegraphics}[2][]{%
      \begin{tikzpicture}
        \node[inner sep=0pt]{\oldincludegraphics[##1]{##2}};
        \node[opacity=0.5, rotate=0, scale=5, text=red] at (0,0) {DRAFT};
      \end{tikzpicture}%
    }%
}{%
    \endgroup
  \end{figure}
}

\usepackage[accepted]{mlsys2025}

\newif\ifcomments
\commentstrue
\ifcomments
    \providecommand{\ion}[1]{{\color{blue}{/* ion: #1 */}}}
\else
    \providecommand{\ion}[1]{}
\fi

\mlsystitlerunning{Unleashing Scalable Context Parallelism for Foundation Models Pre-Training via \textit{FCP}}

\begin{document}

\twocolumn[
\mlsystitle{Unleashing scalable context parallelism for foundation models pre-training via \textit{FCP}}



\mlsyssetsymbol{equal}{*}
\mlsyssetsymbol{intern}{\dag}

\begin{mlsysauthorlist}
\mlsysauthor{Yilong Zhao}{equal,ucb,intern}
\mlsysauthor{Xiaonan Nie}{equal,bd}
\mlsysauthor{Kan Zhu}{uw}
\mlsysauthor{Shuang Ma}{ucd}
\mlsysauthor{Zhichao Lai}{bd}
\mlsysauthor{Hongxiang Hao}{bd}
\mlsysauthor{Yang Zhou}{ucd}
\mlsysauthor{Baris Kasikci}{uw}
\mlsysauthor{Ion Stoica}{ucb}
\end{mlsysauthorlist}

\mlsysaffiliation{ucb}{University of California, Berkeley.}
\mlsysaffiliation{bd}{ByteDance Seed.}
\mlsysaffiliation{uw}{University of Washington.}
\mlsysaffiliation{ucd}{University of California, Davis.}

\mlsyscorrespondingauthor{Xiaonan Nie}{niexiaonan@bytedance.com}

\mlsyskeywords{LLM, Training, Workload Balance, Context Parallelism}

\vskip 0.3in

\begin{abstract}
Context parallelism (CP) has been widely adopted to support the growing context length in foundation model pretraining. 
However, existing designs fail to handle the large variation in sequence length from training datasets, resulting in suboptimal performance. 
These methods often over-shard short sequences, leading to compute inefficiency and excessive communication, or process long and short sequences separately without proper bin-packing, causing workload imbalance.
In this paper, we propose \sys{}, a flexible context parallelism paradigm that shards and schedules sequences at \textit{block-level} granularity, where each sequence is partitioned into fixed-size blocks regardless of its original length. 
Unlike common implementations that rely on rigid communication topologies such as rings,
\sys{} enables arbitrary peer-to-peer communication, allowing flexible placement of \textit{sequence blocks} across workers. 
By bin-packing blocks from both short and long sequences, \sys{} achieves both high compute efficiency and balanced workload distribution. 
Extensive evaluations show that \sys{} attains near-linear scalability on up to $256\times$ NVIDIA GPUs, with $1.13\times$–$2.21\times$ improvement in the attention MFU.
\end{abstract}
]



\printAffiliationsAndNotice{\mlsysEqualContribution \textsuperscript{\dag}Work done during internship at ByteDance Seed. } 

\section{Introduction}
\begin{figure*}
    \centering
    \includegraphics[width=0.9\textwidth]{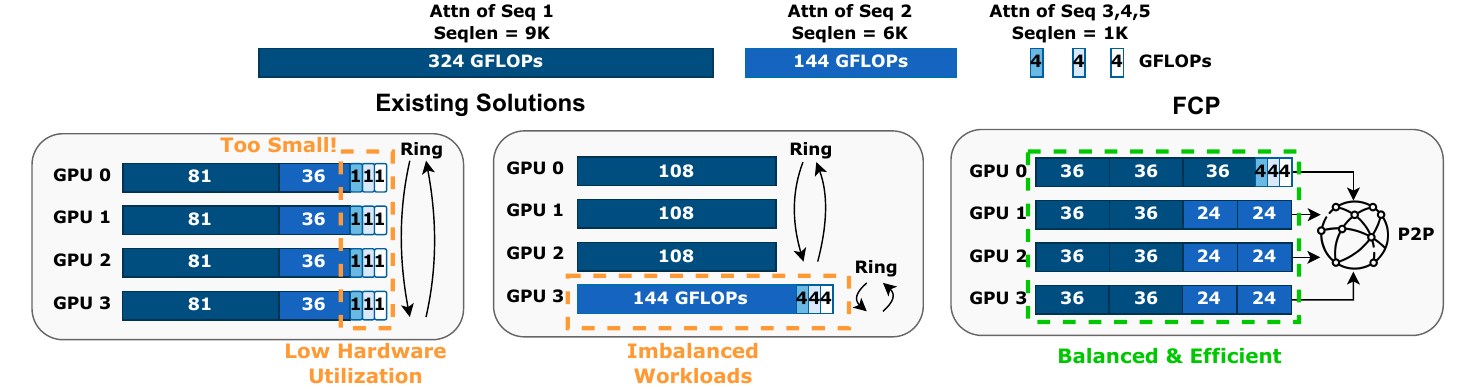}
    \caption{Comparison between \sys{} and existing designs. \textit{(Left)} Compute inefficiency: all sequences are uniformly sharded across GPUs. \textit{(Middle)} Workload imbalance: sequences are grouped by length and assigned to different GPUs. Within each group, ring attention is applied. \textit{(Right)} \sys{} adopts block-grained scheduling with arbitrary peer-to-peer communication.}
    \label{fig:teasor}
\end{figure*}



Training foundation models has become prohibitively expensive, largely because of long-context sequences arising from extended documents and multimodal inputs.
For example, a single 1080p image corresponds to thousands of patch tokens and a one-minute video at 30 fps can expand to millions~\cite{gao2025seedance10exploringboundaries}. Such million-token sequences amplify both memory and compute demands of the \textit{self-attention} module. 
To handle such sequences, \emph{context parallelism} (CP) (e.g., ring attention~\cite{ringattention_arxiv}) was proposed to shard each sequence across several GPUs for attention parallelization, which reduces per-GPU memory usage and scales compute capability.
Combined with model parallelism, CP is expected to keep scaling to larger models and longer sequences~\cite{zheng2022alpaautomatinginterintraoperator, gu2024loongtrainefficienttraininglongsequence}.


However, real-world corpus features high diversity in context lengths, which poses significant complexity and challenges in designing efficient CP algorithms~\cite{Ge_2025,wlb-llm, yang2025contextparallelismscalablemilliontoken}. The first challenge is \textbf{\textit{compute inefficiency}} (\refsec{sec:analysis:hardware-efficiency}). Modern GPUs have specialized matrix-multiplication units (i.e., TensorCores) that need large inputs to saturate~\cite{chen2023punicamultitenantloraserving}. However, short sequences could be over-sharded into tiny ones, which greatly under-utilizes hardware~\cite{wlb-llm}. Even worse, such tiny shards still needs to be transferred across GPUs, causing excessive communication that further decreases compute efficiency. 
The second challenge is \textbf{\textit{workload imbalance}} (\refsec{sec:analysis:workload-balance}). Since attention computation scales quadratically with context length, GPUs handling longer sequences must perform substantially more operations per token than those processing shorter ones. 
As a result, even if each GPU is assigned the same total number of tokens, uneven distribution of context lengths leads to significant imbalance, where some GPUs become overloaded while others remain underutilized, reducing overall cluster efficiency~\cite{wlb-llm}. 





Therefore, achieving optimal CP scheduling requires exploring a huge search space to determine the shard size for each sequence and the mapping of these shards across GPUs. 
In order to reduce the search space, existing designs oversimplify the scheduling problem by compromising either compute efficiency or load balance (\refsec{sec:analysis:existing}), resulting in suboptimal performance. 
For instance, some methods~\cite{ringattention_arxiv,gu2024loongtrainefficienttraininglongsequence,fang2024uspunifiedsequenceparallelism} uniformly split all sequences into the same number of shards, disregarding variation in context length  and causing underutilization of hardware resources. 
Others~\cite{Ge_2025,wlb-llm} separate short and long sequences across different GPUs without mixing them, leading to workload imbalance.

In this paper, we propose \textit{flexible context parallelism}, \sys{}, a new context parallelism paradigm that achieves linear scalability of attention when processing batches of sequences with varying lengths.
The key idea behind \sys{} is fine-grained \textit{block-wise} sharding and a flexible GPU assignment policy.
Specifically, unlike existing designs that follow a predefined pattern (e.g., splitting sequences onto contiguous GPUs as a ring), \sys{} treats each sequence as a series of \textit{blocks} and formulates the scheduling problem as a bin-packing problem to balance the load.
\fig{fig:teasor} shows an example illustrating the differences between \sys{} and existing solutions.



To reduce the search complexity, \sys{} shards each sequence into fixed-size blocks (e.g., $1$K tokens) regardless of its context length, which serves as the basic scheduling and computation unit. 
To maximize compute efficiency, the size of block is set to be large enough to saturate the hardware resources even when processing a single block. Besides, the redundant communication from over-sharding short sequences can be saved. 
To minimize the workload imbalance, \sys{} proposes a workload-aware \cd{} (\refsec{sec:design:block-distributor}). 
In particular, \cd{} estimates the computation and memory usage of each block based on the context length of its source sequence, and applies a variant of Longest-Processing Time (LPT) scheduling algorithm~\cite{chandran2024resultslptnearlineartime} that iteratively assigns blocks to the least-loaded GPUs, achieving near-optimal workload balance.

Despite the theoretical benefit, the efficient implementation of \sys{} in practice raises several challenges.
First, given the flexibility of block assignment, blocks from a single sequence could exist on any GPUs, which requires \ptop{} communication between any two GPUs in the clusters. Therefore, inter-node communication (i.e., InfiniBand or Ethernet) takes a significant fraction of the total time. To reduce this overhead, we need to effectively overlap computation and communication~\cite{wang2024railonlylowcosthighperformancenetwork}. 
In addition, it is non-trivial to determine the optimal communication ordering of each single GPU. For example, random ordering could cause several GPU to pull blocks from the same source GPU simultaneously, which can lead to network congestion.

To address these challenges, \sys{} adopts a block-level pipeline and a congestion-free \cs{} (\refsec{sec:design:communication-planner}).
From a high level, CP essentially comprises three stages: pulling remote blocks, computing corresponding attention blocks, and pushing local blocks to remote. 
To effectively overlap computation with communication, \sys{} decomposes each stage into block-level sub-stages and interleaves them at a block-by-block granularity. 
To avoid congestion across $N$ GPUs at each sub-stage, \sys{} models the communication of blocks as a bipartite graph with $N$ send and $N$ receive nodes, where each edge $i\rightarrow j$ represents a block that transfers from GPU $i$ to $j$. 
From this bipartite representation, \sys{} defines each graph matching as a congestion-free communication round, where each GPU sends and receives at most one block (i.e., one outgoing and one incoming edge).
Therefore, \sys{} iteratively computes maximal matching as a block-level communication plan, which guarantees an optimal order of block transfers (\refsec{sec:design:communication-planner}).



Notably, \sys{} incurs more complex traffic than a ring topology—a trade-off that enables greater flexibility by fine-grained, block-wise scheduling.
Such traffic is guaranteed to overlap with computation, as \cs{} leverages a \textit{performance model} (\refsec{sec:analysis-comm-comp}) to ensure that the computation time is larger than communication at each stage.
We also show that traffic does not become a system bottleneck.

Beyond these performance benefits, \sys{} is designed with modularity in mind, allowing transparent integration with various parallelism approaches, including FSDP~\cite{zhao2023pytorchfsdpexperiencesscaling}, TP~\cite{shoeybi2019megatron}, EP~\cite{deepseekai2024deepseekv2strongeconomicalefficient, nie2023flexmoe}, and SP~\cite{jacobs2023deepspeedulyssesoptimizationsenabling}. 
Instead of requiring users to provide sequences in a specific layout, \sys{} \textit{on the fly} reshuffles the blocks at the beginning of the attention module in each layer and restores them afterward (\refsec{sec:design:reshuffler}).
Therefore, non-attention operations (e.g., computing positional embeddings) can be performed using existing parallelism approaches without modification. Furthermore, the reshuffling is opportunistically overlapped with local attention computation, incurring negligible overhead.

We provide a comprehensive evaluation of \sys{} on the Llama-3-70B model configuration across modern data-center clusters, including up to $256\times$ NVIDIA GPUs (\refsec{sec:eval:hopper-attention}).
We compare \sys{} with \sota{} CP frameworks including ByteScale~\cite{Ge_2025}, WLB-LLM~\cite{wlb-llm}, and RingAttention~\cite{ringattention_arxiv}. We also include a recent open-source project, MagiAttention~\cite{magiattention2025}.
Across three workloads with different context length distributions, \sys{} consistently achieves near-linear scalability, outperforming the baselines by $1.13\times$–$2.21\times$ in attention Model FLOPs Utilization (MFU).
We summarize our contributions as follows:

\begin{itemize}[nosep]
    \item We analyze the inefficiencies of existing CP designs and the challenges in implementing an optimal solution. 
    \item We propose a \textit{block-wise} abstraction that enables fine-grained scheduling for flexible workload partitioning.
    \item We design \sys{} as a modular context parallelism framework that integrates transparently with existing parallelism approaches, including FSDP, TP, EP, and SP. 
    \item We provide an efficient implementation and comprehensive evaluation that demonstrate the generality and feasibility of the proposed method.
\end{itemize}

\section{Background}
\subsection{Pretraining Datasets Feature Long-tailed Length}
\label{sec:bg:workload}

Large language models have recently extended their context window dramatically, from $4$K to $1$M tokens~\cite{apple_foundation,meta_longcontext}. 
Within such long windows, the training datasets exhibit highly \textit{diverse} and \textit{long-tailed} sequence length distributions~\cite{wlb-llm}. 
In addition, the emergence of multi-modality beyond text further amplifies this heterogeneity. 
For example, a minutes-long video sample can be mixed with a short text-only sequence, leading to tens of thousands of tokens difference in context length. 
Such extended context windows and heterogeneous data samples greatly diversify the overall context length distribution. 

\begin{figure}[!t]
    \centering
    \includegraphics[width=\columnwidth]{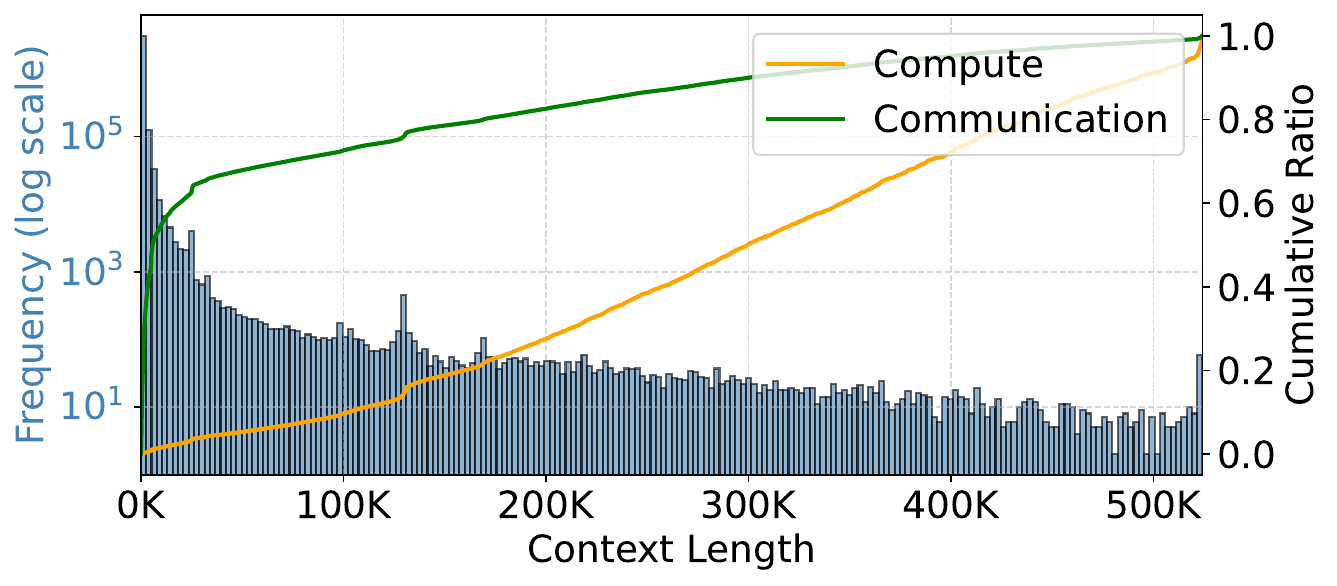}
    \caption{The context length distribution, and cumulative computation and communication ratio from our internal training tasks.}
    \label{fig:input-distribution}
\end{figure}

To illustrate this, we collect the context length distribution from our internal pretraining tasks (with a maximum sequence length of $512$K).
As shown in~\fig{fig:input-distribution}, the context lengths feature long-tailed up to $512$K, approximately following a $\texttt{lognormal}$ distribution. 
We further calculate the total FLOPs and communication volume from each sequence under the naive CP setup and plot the cumulative ratio. 
As shown in~\fig{fig:input-distribution}, both short and long sequences take the same magnitude of computation while short ones dominate communication, necessitating an adaptive parallelization that can efficiently handle diverse datasets.

\subsection{Attention Computation}
Attention is one of the key components in modern foundation models~\cite{vaswani2023attentionneed}.  
Considering a sequence with length $L$, number of attention heads $H$, and head dimension $D$, the attention computation is formulated as:
\[
\mathbf{O} = \text{Softmax}\left(\frac{\mathbf{Q}\mathbf{K}^\top}{\sqrt{D}} \odot \mathbf{M}, \text{dim}=-1\right)\mathbf{V}
\]



Here, $Q$ (query), $K$ (key), $V$ (value), and $O$ (output) are tensors with shape $[H, L, D]$.
Therefore, the overall time complexity of attention is $O(H D L^2)$, while the space complexity is $O(HLD)$. 
This quadratic scaling of computation with respect to $L$ and the linear scaling of memory introduce challenges in balancing computation and memory, especially under batches of diverse $L$.

In this paper, we focus on causal and non-causal attention masks, while leaving irregular patterns for future work. 
Without explicitly written, \textit{attention} in following sections denotes the above-mentioned operations excluding $QKV$ projections. 
As the forward and backward share similar characteristic, while we design and implement both the forward and backward, we discuss the forward for simplicity.

\subsection{Context Parallelism}
To scale pretraining over long-context sequences, attention computation is parallelized across multiple workers by sharding the first two dimensions of input tensors $[H, L, D]$.
For example, since different attention heads compute independently, the head dimension $H$ can be easily split and parallelized. 
This approach is called \textit{sequence parallelism} (SP) in tradition~\cite{jacobs2023deepspeedulyssesoptimizationsenabling, shoeybi2020megatronlmtrainingmultibillionparameter}.
However, as the number of attention heads is typically limited to only a few dozens (e.g., $4$ $KV$ heads in Qwen3-235B~\cite{yang2025qwen3technicalreport}), SP can only scale up to the number of heads, failing to scale up to thousands of workers.

Another common practice for parallelizing attention is splitting along the context dimension $L$, known as \textit{context parallelism} (CP)~\cite{ringattention_arxiv}. 
This is feasible because the attention computation of a single sequence is associative. 
Therefore, intermediate results can be computed by different workers over different slices of $L$ and then reduced to the final outputs.
During CP, each sequence is divided into several \textit{blocks} (i.e., slices of Q/K/V tensors) across workers. 
As blocks from the same sequence have data dependencies, inter-worker communication is introduced.

Assuming $N$ workers, a popular solution, ring attention~\cite{ringattention_arxiv}, uniformly shards each sequence into $N$ blocks, with each worker assigned one block. 
During execution, each worker pulls all $KV$ blocks from one neighbor, computes pulled $KV$ blocks with local $Q$ blocks, and pushes $KV$ blocks to another neighbor, forming a ring topology. 
In practice, ring attention uses double buffering to overlap communication and computation efficiently.
Note that due to memory-efficient attention variants (e.g., group-query attention~\cite{ainslie2023gqatraininggeneralizedmultiquery}), the memory usage of $KV$ tensors is typically smaller than $Q$. Thus, instead of moving $Q$, $KV$ is passed across workers to reduce communication volume. 
In this paper, we mainly focus on efficient CP design due to its potential scalability. 
Meanwhile, SP is orthogonal to CP, and both can be composed together.

\section{Analysis}

\subsection{Compute Efficiency}
\label{sec:analysis:hardware-efficiency}
In context parallelism, attention computation for each sequence is divided into blocks across multiple GPUs, allowing parallel computation. 
Therefore, the overall hardware utilization depends on the compute efficiency of each block $B$. 
However, the block size $\texttt{len}(B)$ cannot be arbitrarily small, as modern accelerators require high \textit{arithmetic intensity} to saturate their compute capacity~\cite{MLSYS2024_5edb57c0}. 
For instance, the latest Hopper GPUs offer up to $989$ TFLOPs of dense BF16 Tensor Core throughput with $4.8$ TB/s of memory bandwidth.
To saturate its compute, each memory-loaded element must be reused $412$ times~\footnote{Assuming FP16 data type, $989/(4.8/2)\approx412$ }, which is further amplified by the complexity of attention kernels~\cite{dao2022flashattention}.

\begin{figure}[!t]
    \centering
    \includegraphics[width=\columnwidth]{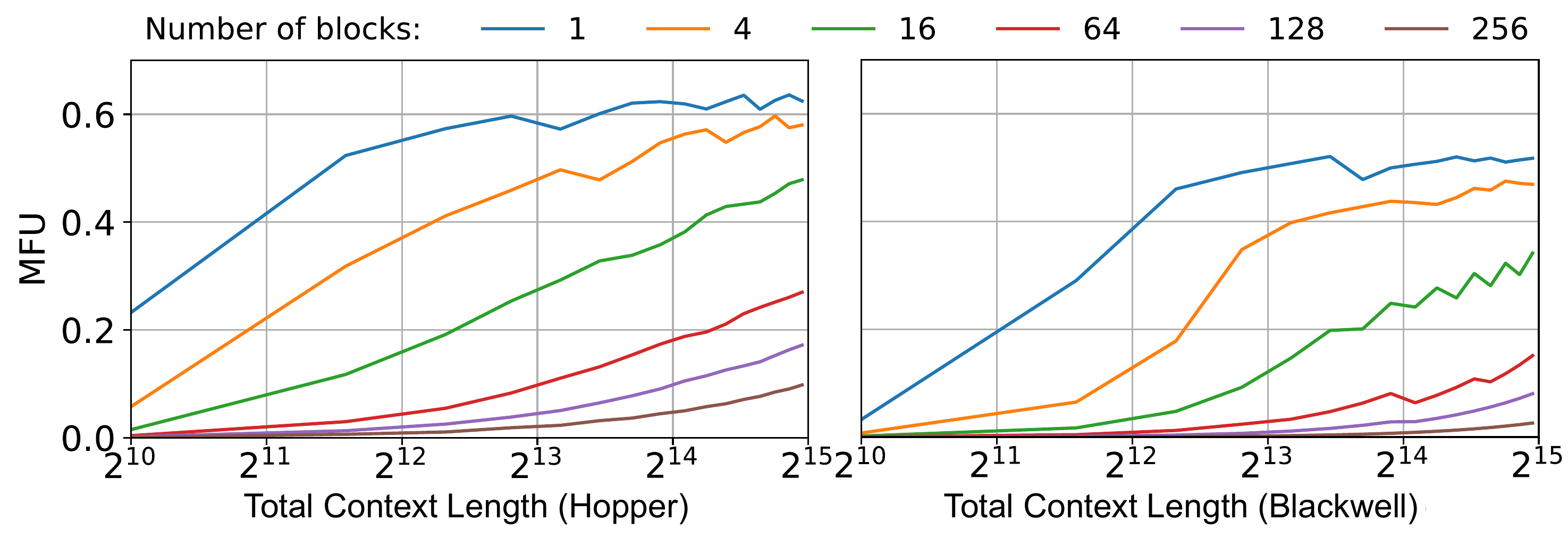}
    \caption{MFU of attention on different hardware, which is profiled with $8$ $KV$ heads, $64$ $QO$ heads, and a head dimension of $128$. We vary the total context length and the number of blocks that compose this total number of tokens. Results showcase that sharding sequences into fine-grained blocks greatly hurt MFU.\footnotemark{}} 
    \label{fig:compute-efficiency}
    \vspace{-4mm}
\end{figure}

To demonstrate this, we measure the model FLOPs utilization (MFU) of state-of-the-art $\texttt{variable-length}$ attention implementations, namely FA3~\cite{shah2024flashattention3fastaccurateattention} on Hopper GPUs and FA4 on Blackwell GPUs.
By varying total context length and the number of sharded blocks, we observe that smaller $\texttt{len}(B)$ (i.e., larger number of blocks under the same context length ) drastically reduces MFU. 
As shown in~\fig{fig:compute-efficiency}, MFU remains low when $\texttt{len}(B)<2$K and begins to saturate beyond $4$K. 
For example, with a total context length of $32$K from $64$ blocks, each block has $512$ tokens, leading to $25$\% utilization.

Furthermore, sharding small sequences introduces unnecessary communication, which further decreases compute efficiency.  
For example, sequences with lengths smaller than $4$K can easily fit into a single worker without incurring any communication. 
As shown in~\fig{fig:input-distribution}, such sequences account for about $50\%$ of the total communication volume in vanilla context parallelism, which can be saved by not sharding with a flexible parallelization strategy.
Consequently, the actual computation time needs to be modeled as $f(B)$, which considers both the total compute amount and the compute efficiency determined by block size.

\footnotetext{All results were profiled on UC Berkeley's computing cluster.}

\subsection{Workload Balance}
\label{sec:analysis:workload-balance}
In addition to single-GPU compute efficiency, the workload assignment $M: B \rightarrow \texttt{worker}_{id}$ significantly affects the overall MFU. 
A single straggler with heavier workloads can leave all other workers idle, reducing cluster utilization. 
The workload balance needs consider three types of resource required by executing a single sequence shard $B$:

\MyPara{Memory.}
The memory usage of worker $i$ equals the sum of $\texttt{len}(B)$ where $M(B)=i$. Memory balance is critical to the end-to-end MFU, as it influences the compute balance of \textit{non-attention} modules such as FFN. 
It also constrains the global batch size, as a high memory usage potentially leads to out-of-memory during training. 
Therefore, an ideal solution should enforce strict limits on the total number of tokens assigned to each worker.

\begin{figure}[t]
    \centering
    \includegraphics[width=\columnwidth]{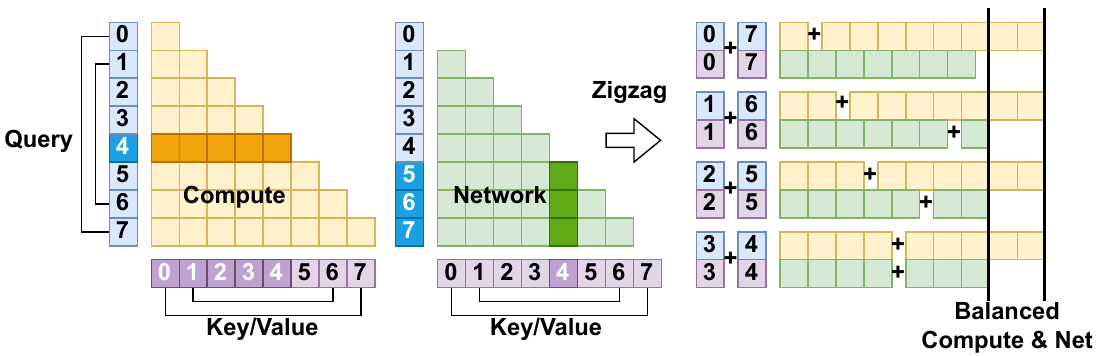}
    \vspace{-4mm}
    \caption{Zig-Zag packing of an $8$-tokens sequence for intra-sequence computation and communication balance under causal mask. The computation and communication volume of a block depends on its position within the sequence. For example, $4$-th $Q$ needs to compute with $5$ KV blocks, while $4$-th $KV$ is transferred $3$ times to the subsequent $Q$ blocks. By packing $i$-th block with $(2N-i)$-th block, both resources can be perfectly balanced.}
    \label{fig:ana_strip}
\end{figure}

\MyPara{Compute.}
Existing studies have extensively explored balancing the computation of a \textit{single sequence} (i.e., intra-sequence) under both causal and non-causal masks~\cite{Brandon2023StripedAF, li2024distflashattndistributedmemoryefficientattention}. 
For example, as shown in~\fig{fig:ana_strip}, ring attention assigns $B$ to workers in a Zig-Zag order, ensuring that each worker performs the same amount of computation. 
However, balancing both compute and memory across \textit{multiple sequences} (i.e., inter-sequence) with different context lengths is non-trivial, as computation grows quadratically with the context length and $\texttt{len}(B)$ scales linearly.
Therefore, achieving balanced overall workload requires considering inter-sequence length variation.

\MyPara{Communication.}
The communication volume of each worker is also crucial, as a network hotspot causes congestion and reduces the overall bandwidth utilization. 
Fortunately, communication balance can be reduced to computation balance for both causal and non-causal masks. 
As shown in~\fig{fig:ana_strip}, although the communication volume of $B$ depends on its position within the sequence, the Zig-Zag assignment ensures that communication volume is proportional to computation. 
In this paper, unless otherwise specified, Zig-Zag ordering is applied for causal masks, simplifying communication balance into computation balance.

\subsection{Optimal Context Parallelism Scheduling}
\label{sec:analysis:problem-formulation}

Given a set of $\tau$ input sequences $S = \{s_1, ..., s_{\tau}\}$ with different lengths, and a set of $N$ workers $W = \{w_1, ..., w_N\}$, each context parallelism scheduling is defined by a sharding function $G$ and an assignment function $M$.
The sharding function $G$ determines how many blocks are generated from each sequence and how large each block is, that is, $G: s_i \rightarrow \{B_{i1}, ..., B_{ik_i}\}$ where $k_i$ denotes the number of blocks from $s_i$. 
The assignment function $M$ maps each block to one of the workers, $M: B \rightarrow w_i$.

For the $i$-th worker $w_i$, its computation load $\texttt{Comp}(w_i)$ can be calculated as $\texttt{Comp}(w_i) = \sum_j f(B_j) \cdot I_{M(B_j) = w_i}$, where $f(\cdot)$ considers both FLOPs and compute efficiency defined in~\refsec{sec:analysis:hardware-efficiency}, and $I$ is the indicator function. 
With a network overlapping factor $\eta_i \ge 1$, the total time for $w_i$ is $\eta_i \cdot \texttt{Comp}(w_i)$, where perfect overlap of computation and communication is achieved when $\eta = 1$. 
Note that constants are ignored for simplicity.
Considering all workers, the end-to-end time is $T = \max_{i \in N} (\eta_i \cdot \texttt{Comp}(w_i))$. 
Therefore, the optimal context parallelism scheduling $(G, M)$ is given by $(G^*, M^*) = \arg\min_{G, M} (T)$, which achieves both high compute efficiency and balanced workloads.

However, solving this optimization problem is apparently $\texttt{NP-complete}$, with an exponential computational complexity. 
It is thus infeasible to obtain a meaningful solution within practical time. Therefore, all existing approaches approximate the optimal scheduling by reducing the search space through simplified assumptions.

\subsection{Existing Context Parallelism Fails to Approach Optimal Scheduling}
\label{sec:analysis:existing}
To derive a practical strategy within a reasonable time, existing solutions simplify either $G$ or $M$ through \textit{human-crafted}, \textit{rule-based} heuristics. 
Such simplifications lead to designs that overfit to specific workloads and lack adaptability to the diverse real-world corpus as discussed in~\refsec{sec:bg:workload}.

\begin{figure}
    \centering
    \includegraphics[width=\columnwidth]{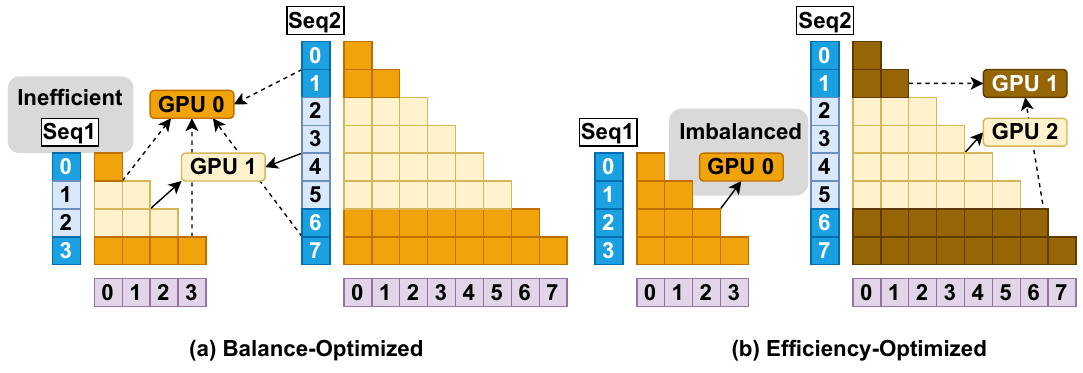}
    \vspace{-4mm}
    \caption{Illustration of two kinds of existing CP designs. (a) balance-optimized: each sequence is sharded into $2\times2=4$ blocks and packed with Zig-Zag ordering, thus achieving perfect balance. (b) efficiency-optimized: sequences are spatially partitioned into different workers based on the context length.}
    \label{fig:existing-pattern}
\end{figure}
\MyPara{Balance-optimized.}  
Ring attention~\cite{ringattention_arxiv} oversimplifies $G$ and fails to consider compute efficiency. It shards each sequence into $2N$ blocks, where $N$ is the number of workers, and assigns two blocks from each sequence to every worker (with the $i$-th and $(2N-i)$-th blocks). 
As shown in~\fig{fig:existing-pattern} (a), when combined with Zig-Zag ordering, this heuristic can achieve perfect workload balance across workers. 
However, it fixes $\forall i \in \tau, k_i = 2N$, without considering both the context length and compute efficiency $f(\cdot)$. 
As a result, short sequences are over-sharded into very small blocks with $len(B) < 2$K, causing severe MFU degradation as discussed in~\refsec{sec:analysis:hardware-efficiency}.

\begin{figure*}
    \centering
    \includegraphics[width=0.85\textwidth]{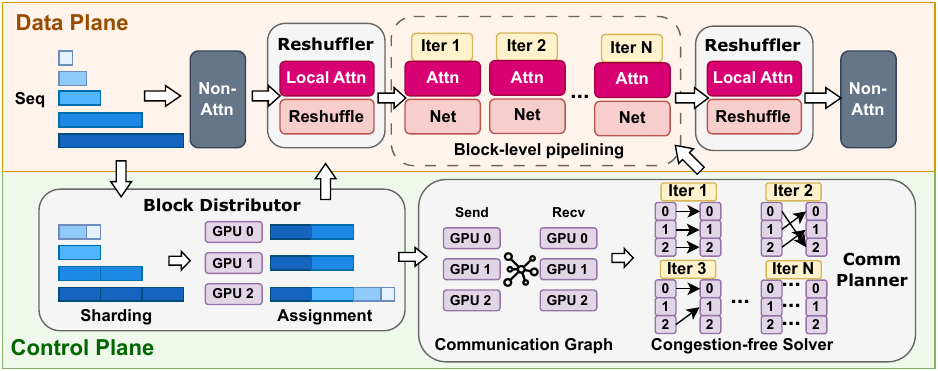}
    \caption{\sys{} System Overview. 
    }
    \label{fig:design-overview}
\end{figure*}

\MyPara{Efficiency-optimized.}  
To improve compute efficiency, other works~\cite{Ge_2025} make $G$ context-length-aware instead of applying constant $k$. 
They cluster sequences into groups according to their context length and allocate different numbers of workers to each subgroup proportionally to the context length. 
As shown in~\fig{fig:existing-pattern} (b), ring attention is applied within each subgroup.
However, these designs oversimplify $M$ by isolating subgroups on separate workers without resource sharing. 
Consequently, outlier sequences with extreme context length (which are common as shown in~\refsec{sec:bg:workload}) can severely disturb workload balance. 
For example, a $64$K sequence requires $256\times$ more computation than a 4K one but is only given $16\times$ more compute resources, resulting in $16\times$ higher per-worker computation. 
Such a rigid constraint on the search space of $M$ greatly hurts workload balance thus the overall MFU.

\MyPara{Switch between two.}  
Other studies~\cite{wlb-llm} identify that existing approaches perform poorly on certain workloads. 
Therefore, they propose adaptively switching between the two aforementioned methods through a profile-guided performance estimator, to achieve best of two. 
However, these solutions do not expand the limited search space, which fundamentally reflects a trade-off between compute efficiency and workload balance. As a result, they fail to achieve both objectives simultaneously.

\subsection{Ring-Based Assignment Policy: A Gilded Cage}
\label{sec:analysis-comm-comp}

Despite different implementations of the sharding policy $G$, existing CP designs share the same principle for the assignment policy $M$ (\refsec{sec:analysis:existing}). 
They all adopt a ring-based topology, either as a monolithic ring or as heterogeneous sub-rings.
This is because a ring-based topology provides a simple and symmetric communication pattern. 
Each worker only exchanges data with its neighboring workers, which enables high network bandwidth utilization and better communication–computation overlap (i.e., lower $\eta$).

However, we argue that the ring topology is not the only way to achieve efficient communication. 
Moreover, its rigid constraint requires each sequence to be sharded symmetrically across a fixed ring of workers, which greatly limits the search space, making CP scheduling fundamentally sub-optimal.
To demonstrate this, we analyze the computation and communication demands of a block $B$ under various worker types and network configurations. 
We vary $len(B)$ and compute the network bandwidth required for the communication time of $B$ to match its computation time.

For example, with the latest Hopper GPUs and a $50$GB/s ConnectX-7 InfiniBand network, only $22$GB/s ($44$\% of the line rate) is required to fully overlap communication with computation (i.e., $\eta=1$). 
Furthermore, increasing the block size $len(B)$ reduces the bandwidth requirement, because computation within each block scales quadratically with the block size, whereas communication scales linearly. This enables a flexible trade-off between scheduling granularity and communication efficiency.

\subsection{\sys{}: Enabling Flexible Context Parallelism with Block-wise Sharding and Assignment}

\MyPara{Opportunity.} 
In this paper, we aim to remove the ring-topology constraint in the search space of the assignment policy $M$. 
By enabling arbitrary peer-to-peer communication, each sequence block $B$ can be placed arbitrarily on any worker. 
This flexibility allows the system to achieve ideal workload balance under various context length distributions.

\MyPara{Challenges.} 
However, realizing this flexibility raises several challenges:
(1) how to maintain high compute efficiency $f(\cdot)$ for each single batch $B$. 
(2) how to balance workloads across workers in terms of memory, computation, and communication. 
(3) how to enable efficient communication (i.e., $\eta=1$) to avoid system bottlenecks.

\section{Design}
\label{sec:design}
As shown in~\fig{fig:design-overview}, \sys{} consists of three main components: \textit{\cd{}} (\refsec{sec:design:block-distributor}), \textit{\cs{}} (\refsec{sec:design:communication-planner}), and \textit{\tr{}} (\refsec{sec:design:reshuffler}).
Given a set of sequences from a training batch, \textit{\cd{}} determines how to shard these sequences and assign the blocks to workers. 
It takes into account both compute efficiency and load balance.
Based on the block assignment, \textit{\cs{}} builds a communication bipartite graph according to the dependencies among blocks, and derives an optimal communication plan using a congestion-free solver. 
These plans are executed through block-level pipelining, which enables overlap between communication and computation.

To make \sys{} transparent to existing frameworks, \textit{\tr{}} reshuffles the user-provided sequence layout into the workload-aware layout required by \sys{} when entering the attention module, and restores the original layout afterward. 
To minimize overhead, the reshuffling is integrated into the block-level pipeline and overlapped with local attention computation whenever possible.

\subsection{\CD{}}
\label{sec:design:block-distributor}
\MyPara{Sharding Policy.}
Given a set of input sequences, \cd{} first divides each sequence into a list of \textit{fixed-size blocks}, for the following two main advantages:
First, it greatly reduces the search space of optimal scheduling without losing expressiveness. 
It adapts to various context lengths, producing more blocks for longer sequences and fewer for shorter ones. 
Second, since each block shares the same communication and computation volume per execution with another block (although the total number of executions may differ), the block-wise abstraction allows systematic modeling and scheduling at the block level (\refsec{sec:design:communication-planner}).
For sequences with context length shorter than block size, \sys{} packs them into minimal number of blocks and adopts the $\texttt{varlen}$ API of the attention kernel for computation.

\begin{figure}
    \centering
    \includegraphics[width=\columnwidth]{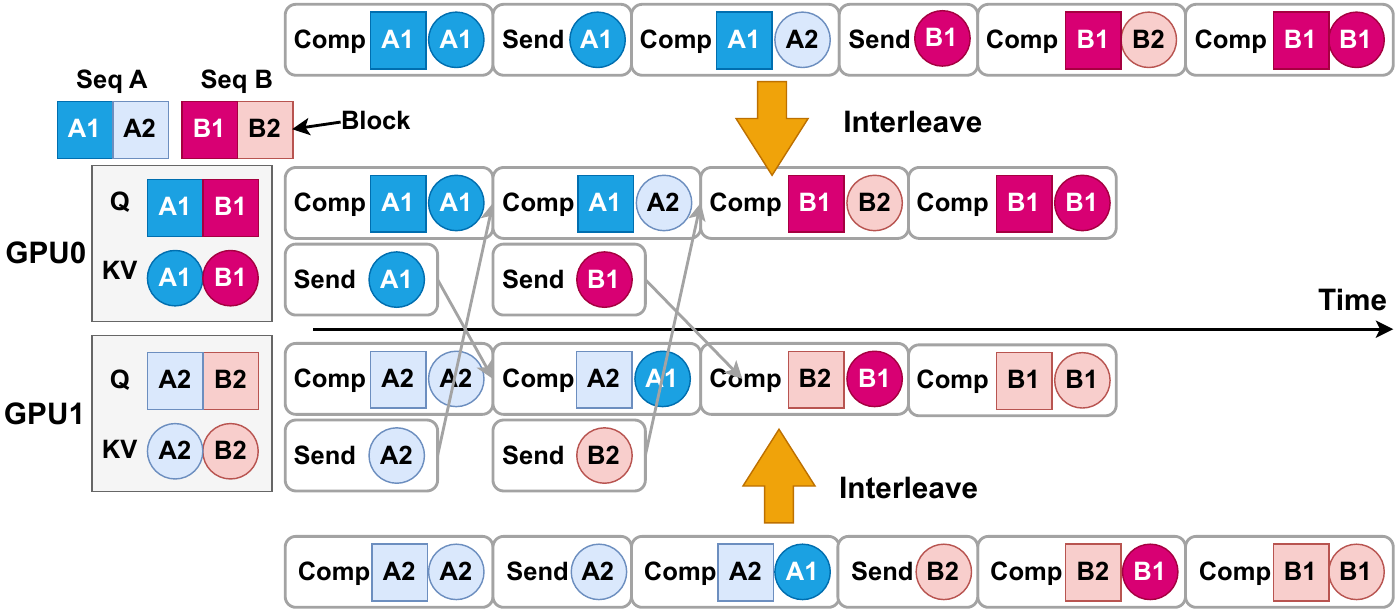}
    \caption{Example of block-level pipelining for efficient computation and communication overlap. \sys{} decomposes end-to-end execution into computation and communication of blocks, which are executed block-by-block in an interleaving way.}
    \label{fig:overlap-pipeline}
\end{figure}

To maximize compute efficiency, the block size $len(B)$ is determined by both hardware and model configurations, ensuring it can fully utilize $f(\cdot)$ as discussed in~\refsec{sec:analysis:hardware-efficiency}. 
Moreover, $len(B)$ also depends on the network configuration to enable effective communication and computation overlap as shown in~\refsec{sec:analysis-comm-comp}. 
Therefore, $len(B)$ is a trade-off between scheduling granularity and runtime efficiency. 
We provide a sensitivity study on different choices of $len(B)$ in~\refsec{sec:eval:sensitivity-test}.

\MyPara{Assignment Policy.}
After sharding sequences into blocks, \cd{} distributes these blocks across workers. 
The objective is to minimize the maximum computation workload among all workers, subject to the constraint of maximal memory usage per worker. 
\sys{} employs a variant of the well-known Longest Processing Time (LPT) scheduling algorithm, which greedily assigns each block to the least loaded worker. 
We describe the policy in detail in Appendix~\ref{sec-appendix-assignment-policy}.
Given $K$ blocks and $N$ GPUs, the time complexity is $O(K\log N)$, which incurs negligible latency.

\begin{figure*}
    \centering
    \includegraphics[width=0.95\textwidth]{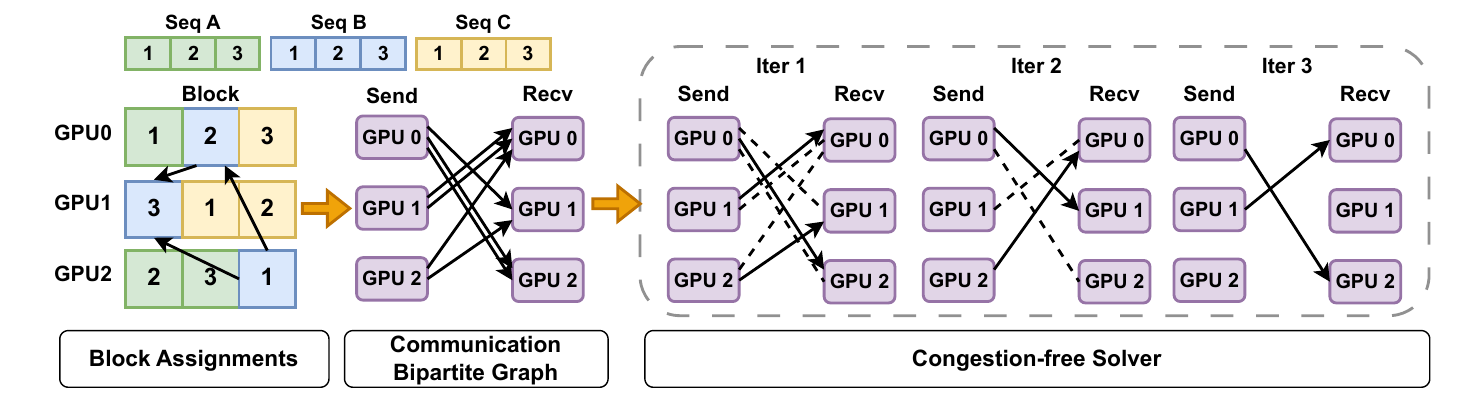}
    \caption{Example of the congestion-free solver over three sequences with causal mask. Given the block assignments from \cd{}, \cs{} constructs a bipartite graph based on the data dependency across GPUs. For example, $1$-st block from sequence $B$ are transferred from GPU $2$ to GPU $0$ and $1$, adding edges $2\rightarrow1$ and $2\rightarrow0$. The solver then calculates the maximal matching of the bipartite graph with minimum iterations, providing an optimal congestion-free communication ordering.}
    \label{fig:comm}
\end{figure*}


\subsection{\CS{}}
\label{sec:design:communication-planner}
After sharding and assignments, the data flow (i.e., data dependency among blocks between workers) is determined. 
However, due to arbitrary peer-to-peer communication, naive communication without overlapping computation introduces huge overhead. 
To achieve efficient communication, \sys{} introduces three techniques: block-level pipelining, a congestion-free solver, and a bottom-up coalescer. 

\MyPara{Block-level pipelining.} 
\sys{} divides both computation and communication across workers into block-grained sub-stages. 
This design allows flexible and predictable scheduling. 
Specifically, as CP consists of three main stages including pulling remote blocks, computing attention blocks, and pushing local blocks, \sys{} decomposes each stage into fine-grained block-level sub-stages and interleaves them block by block. 
For example, as shown in~\fig{fig:overlap-pipeline}, each worker pulls or pushes only one block at a time, while simultaneously computing the previous block, thus overlapping communication and computation effectively.

\MyPara{Congestion-free solver.} 
The ordering of pipelining sub-stages is crucial. 
A random ordering may cause multiple workers to pull blocks from the same worker, resulting in network congestion.
To ensure congestion-free communication, \sys{} models the data flow among $N$ workers as an \textit{undirected bipartite graph}. 
The graph consists of $N$ send nodes $S$ and $N$ receive nodes $R$, where an edge $S(i) \rightarrow R(j)$ indicates that a block must be transferred from the $i$-th to the $j$-th worker. 
As shown in~\fig{fig:comm}, once the block assignments are determined, the bipartite graph can be directly constructed based on the data dependencies. 
The maximal degree $\Delta$ of the bipartite represents the workload balance defined by \cd{}. 

We then show that the optimal number of congestion-free sub-stages is $\Delta$, and provide a practical solver with polynomial-time complexity. 
A single-stage congestion-free plan means each worker sends at most one block, receives at most one block, and computes at most one block. 
Therefore, for all $i\in N$, $\texttt{deg}(S(i))\le 1$ and $\texttt{deg}(R(i))\le 1$. 
This is exactly the definition of a matching in the corresponding bipartite graph. Thus, we have Lemma~\ref{lem:cf-matching}.

\begin{lemma}
\label{lem:cf-matching}
A single-stage congestion-free communication is exactly a matching on the bipartite communication graph.
\end{lemma}

In each sub-stage, a worker transfers at most one block. 
The worker with degree of $\Delta$ needs at least $\Delta$ sub-stages to finish. 
Hence, the minimal number of congestion-free sub-stages is at least $\Delta$. 
Thus, we have Lemma~\ref{lem:num-matching}.
We then construct a solution that partitions all edges into $\Delta$ \textit{disjoint matchings}. 
We detailed the construction in Appendix~\ref{sec-appendix-solver}. 
Therefore, combining with Lemma~\ref{lem:cf-matching} and ~\ref{lem:num-matching}, the optimal number of congestion-free sub-stages equals $\Delta$.

\begin{lemma}
\label{lem:num-matching}
A bipartite with maximal degree $\Delta$ is partitioned into at least $\Delta$ disjoint matchings.
\end{lemma}

Assuming an edge set $E$ and a vertex set $V$, the communication plans can be calculated with $O(|V|^{1/2}|E|)=O(\hat{N}^{2.5})$ time complexity via Hopcroft–Karp algorithm~\cite{Hopcroft}, where $\hat{N}$ is the CP group size (typically in the hundreds).
Given the sharding and assignments of the sequences, \sys{} will first construct the corresponding bipartite, which is then partitioned into a set of matchings. 
At each sub-stage, \sys{} follows one matching to determine the network flow, while the communication is overlapped with the computation of one block. 
Moreover, the solver runs once per batch, which can be either overlapped or pre-computed. 
Since total GPUs are partitioned into independent CP groups and different training batches are independent, the matchings are massively parallelizable across CPUs, completing within seconds at the scale of hundreds of workers.
We illustrate the process in~\fig{fig:comm}.

\MyPara{Bottom-up coalescer.}
Furthermore, the fine-grained block-wise sub-stages can be coalesced into coarse-grained stages without compromising congestion-free communication. 
For example, assuming a coalesce degree of $4$, the computation and communication of four consecutive sub-stages can be merged into a larger stage. 
In each sub-stage, every worker sends, receives, and computes at most one block. 
After coalescing, each worker sends $4$, receives $4$, and computes $4$ blocks in a single stage without incurring a network hotspot.
This merging improves kernel efficiency by increasing block sizes of execution (\refsec{sec:analysis:hardware-efficiency}).
Therefore, the coalescer essentially decouples the scheduling granularity (i.e., block size) from the execution granularity (i.e., coalesced block size), enabling more flexible scheduling of \sys{}.

\subsection{\TR{}}
\label{sec:design:reshuffler}
Even with efficient implementation, CP is hard to deploy in real-world systems. 
This is because it requires intrusive modifications to existing frameworks. 
For example, aligning the sequence layout with desired assignments requires modifying sequence-dependent components, including the dataloader~\cite{Ge_2025} and the positional embedding (e.g., RoPE)~\cite{magiattention2025}.

Therefore, making \sys{} transparent is important for deployment. 
A strawman solution is to insert two all-to-all communications before and after the attention module. 
These operations on-the-fly reshuffle the user-provided sequence layout into the workload-aware layout required by CP.
However, such standalone all-to-all operations can introduce additional overhead. 
Fortunately, the communication of all-to-all can be overlapped with computation. 
This is because the total communication volume is bounded by the total context length, as each sequence only has one copy. Meanwhile, the total computation grows quadratically. 

To achieve overlap between reshuffling and computation, \sys{} leverages the fine-grained block-level pipeline (\refsec{sec:design:communication-planner}). 
It reorders local computation (i.e., computation that does not depend on remote blocks) to the beginning and end of the pipeline. 
When entering or exiting attention, \sys{} first launches local computation before the reshuffling, enabling effective overlap between communication and computation. 

\section{Implementation}


\MyPara{Implementation details.} 
We implement \sys{} with $4$K lines of Python code, with minor modifications on FlashAttention3~\cite{shah2024flashattention3fastaccurateattention} kernels. 
To prevent interference between computation and communication, we employ CUDA Green Context~\cite{cui2025optimizingsloorientedllmserving} to spatially partition GPU SMs into communication and computation ones. 
Since the communication operations are pure data transport without reduction, the network (e.g., $50$~GB/s InfiniBand) rather than HBM ($2$~TB/s) is the bottleneck, so a small number of SMs suffices to saturate the link bandwidth.
We set the number of communication SMs to the minimum feasible value following the CUDA driver API~\cite{cuda_green_ctx}.
We specify $\texttt{sm\_margin}$ in the attention kernel to mitigate wave quantization~\cite{osama2023streamkworkcentricparalleldecomposition}. 
For communication, we use the group peer-to-peer primitives of NCCL and make the communication pattern aware of the network topology. 
For the railed-optimized network topology~\cite{wang2024railonlylowcosthighperformancenetwork}, we enable Ulysess~\cite{jacobs2023deepspeedulyssesoptimizationsenabling} with \sys{} to avoid two-hop communication. 
For block-level pipelining (\refsec{sec:design:communication-planner}), we implement a multi-buffer pipeline that launches several communication operation concurrently, reducing pipeline bubbles. 
In our evaluation, we use three buffers. 

\section{Evaluation}

\subsection{Evaluation Setups}
\label{sec:eval:setup}

\MyPara{Hardware.} 
We conduct the evaluation on two computing clusters from industry and academia, respectively. Across these clusters, we use two GPU types, denoted as GPU-X and GPU-Y. Their computation-to-communication ratios, measured as BFloat16 TensorOp throughput divided by network bandwidth, are summarized in Table~\ref{tab:eval-hardware}. To respect confidentiality constraints, we anonymize certain system details, including the exact GPU models and the cluster scale.

\begin{table}[h]
\centering
\small
\vspace{-4mm}
\caption{Computation-to-communication ratios of GPU-X and GPU-Y.}
\label{tab:eval-hardware}
\begin{tabular}{lcc}
\toprule
 & \textbf{GPU-X} & \textbf{GPU-Y} \\
\midrule
\textbf{Comp/Comm} & 5920 & 2500 \\
\bottomrule
\end{tabular}
\vspace{-2mm}
\end{table}

\MyPara{Model and workload.} 
We adopt the model configuration from Llama-3-70B~\cite{grattafiori2024llama3herdmodels}, with $8$ $KV$ heads, $64$ $QO$ heads and a head dimention of $128$. 
We randomly sample sequences from our training traces as shown in~\fig{fig:input-distribution}, with a maximum sequence length of $512$K. 
We provide results with more datasets in Appendix~\ref{sec-appendix-more-workload}.
We set the number of tokens per GPU as $32$K, which is common practice in large-scale pretraining. We also test the performance under different number of per-GPU tokens in~\refsec{sec:eval:sensitivity-test}.
We apply the causal attention mask for all sequences.

\MyPara{Baselines.} 
We compare \sys{} with the following \sota{} CP designs: 
\ding{172} \textit{Ring Attention}: Balance-optimized design.
\ding{173} \textit{ByteScale}: Compute-efficiency optimized design that dynamically partitions short and long sequences into different GPUs~\cite{Ge_2025}.
\ding{174} \textit{WLB-LLM}: adaptive switch between balance- and efficiency-optimized baselines based on an online performance estimator. 
Besides these well-established literature, we also include a concurrent work from an open-sourced project, MagiAttention (\ding{175})~\cite{magiattention2025}, to provide comprehensive comparison. 
We provide detailed configuration of baselines in Appendix~\ref{sec-appendix-baselines}.

\MyPara{Hyper-parameters.}
We use a $4$K block size for both GPU-X and GPU-Y, while we also conduct a sensitivity test over various block sizes in~\refsec{sec:eval:sensitivity-test}. 
We use a coalesce degree of $16$ by default for better efficiency (\refsec{sec:design:communication-planner}).
We assign $6$ and $8$ SMs for communication operators on GPU-X and GPU-Y.

\begin{figure}[t]
    \centering
    \includegraphics[width=\columnwidth]{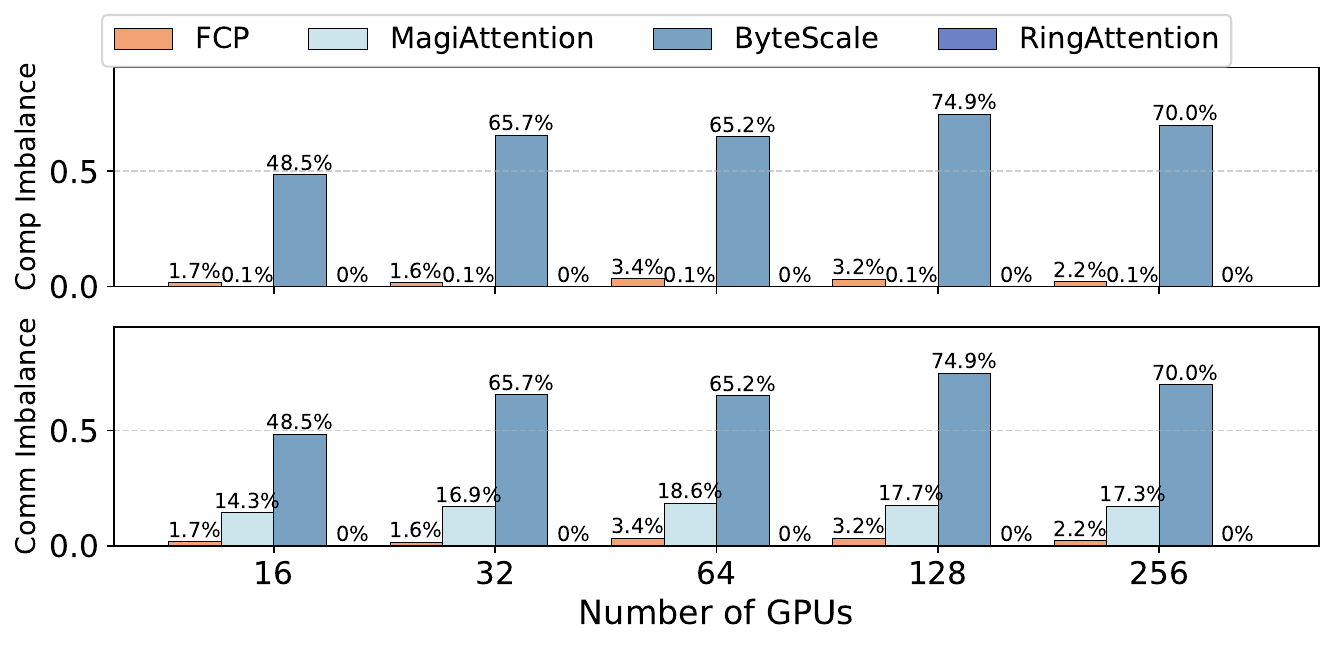}
    \vspace{-6mm}
    \caption{Computation (upper) and communication (lower) imbalance ratio when scaling the number of GPUs.}
    \label{fig:eval-balance}
\end{figure}
\begin{figure}[t]
    \centering
    \includegraphics[width=\columnwidth]{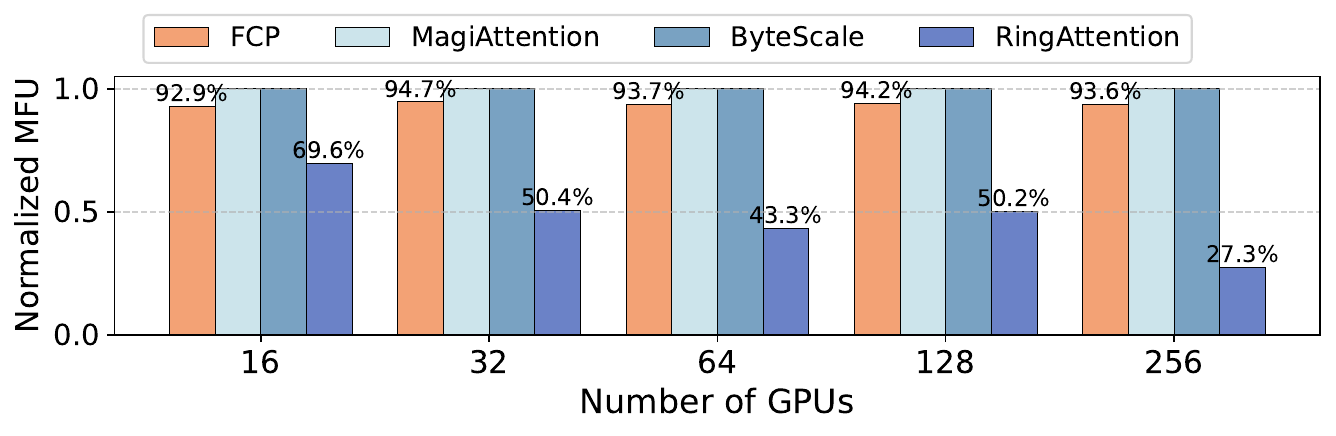}
    \vspace{-6mm}
    \caption{Normalized attention MFU with perfect load balance.}
    \label{fig:eval-compute-efficiency}
    \vspace{-2mm}
\end{figure}

\subsection{Attention Workload Balance}
\label{sec:eval:balance}
We evaluate the workload balance (i.e., computation and communication volume) by accumulating the FLOPs and network traffic of each GPU.
As the GPU with heaviest workload bottlenecking the cluster, we define the \textit{imbalance ratio} as $(\texttt{max}(load)-\texttt{mean}(load))/\texttt{max}(load)$. 
As shown in~\fig{fig:eval-balance}, \sys{} consistently achieves less than $5$\% workload imbalance, owing to its fine-grained block-level block assignments. 
In contrast, as MagiAttention only optimizes for computation, it incurs up to $17$\% communication volume imbalance, leading to sub-optimal communication. 
Besides, as ByteScale spatially partitions sequences based on their context length $L$, the $O(L^2)$ computation is only assigned with $O(L)$ GPUs, causing up to $70$\% imbalance.

\subsection{Attention Compute Efficiency}
\label{sec:eval:hopper-attention}
To measure the compute efficiency excluding the effect of workload imbalance, we assume all context lengths equal to the average length of~\fig{fig:input-distribution}. 
We compute the attention MFU as the ratio of total computation to the total attention time, normalized by the MFU of single-GPU FlashAttention.
As shown in~\fig{fig:eval-compute-efficiency}, \sys{} consistently achieves more than $90$\% MFU, with the minimal degradation due to SMs specialized for communication. 
In contrast, Ring Attention greatly decreases MFU due to the over-sharding on short sequences, increasing computation and communication time.

\begin{figure}[t]
    \centering
    \includegraphics[width=\columnwidth]{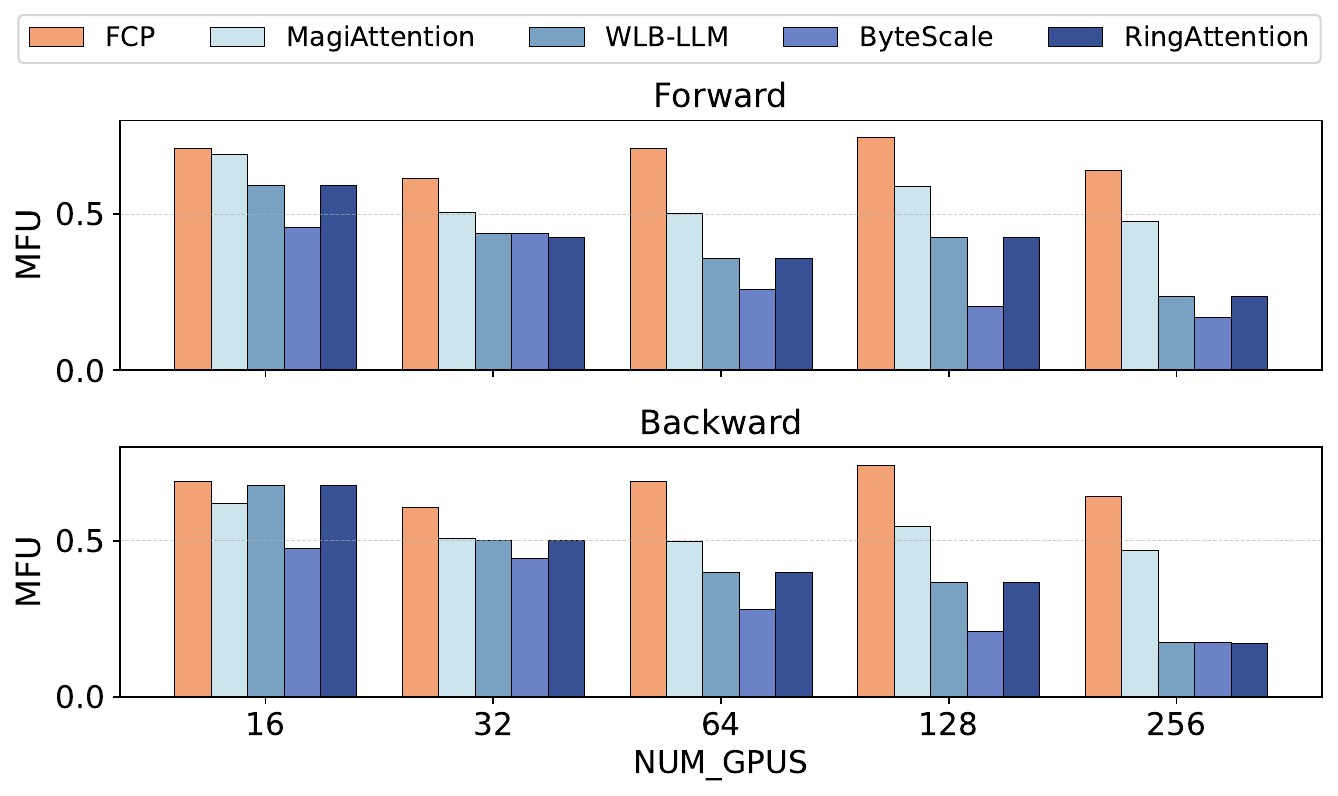}
    \vspace{-4mm}
    \caption{\textit{Weak-scaling} of module-level attention MFU on real-world dataset. The number of tokens per GPU is fixed at $32$K.}
    \label{fig:eval-mfu}
    \vspace{-4mm}
\end{figure}

\begin{figure}[t]
    \centering
    \includegraphics[width=0.9\columnwidth]{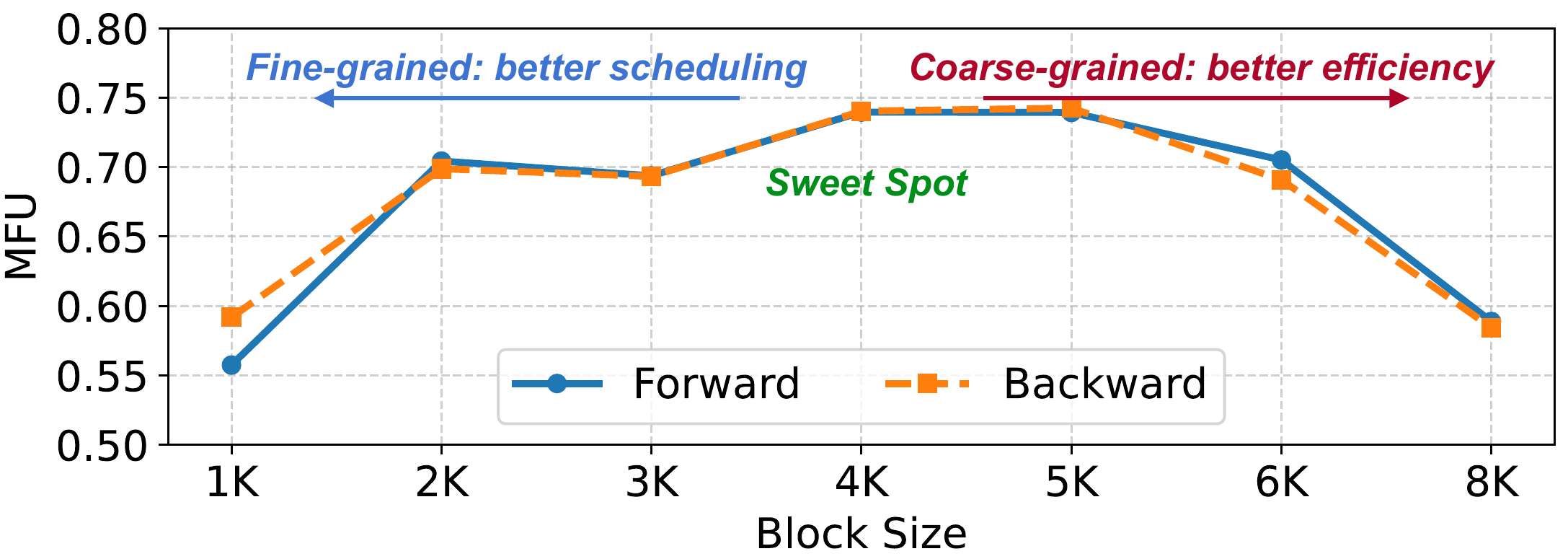}
    \vspace{-4mm}
    \caption{Sensitivity test of block sizes on $128\times$ GPU-X.}
    \label{fig:eval-sensitivity-block-size}
\end{figure}

\begin{table}[b]
\centering
\small
\vspace{-2mm}
\caption{Ablation study of MFU on $128\times$ GPU-X.}
\label{tab:eval-ablation}
\begin{tabular}{lccccc}
\toprule
 & Base & \#1 & \#2 & \#3 & \#4 \\
\midrule
\textbf{Fwd} & 0.29 & \makecell{0.48\\ (64\%$\uparrow$)} & \makecell{0.62\\ (29\%$\uparrow$)} & \makecell{0.70\\ (10\%$\uparrow$)} & \makecell{0.75\\ (7\%$\uparrow$)} \\
\textbf{Bwd} & 0.37 & \makecell{0.46\\ (24\%$\uparrow$)} & \makecell{0.59\\ (28\%$\uparrow$)} & \makecell{0.69\\ (17\%$\uparrow$)} & \makecell{0.74\\ (7\%$\uparrow$)} \\
\bottomrule
\end{tabular}
\vspace{-2mm}
\end{table}

\subsection{Scaling Test of Module-level MFU}
We further measure the end-to-end attention-module MFU at the cluster level, considering for both inter-GPU workload imbalance and single GPU compute efficiency. 
The module-level MFU is defined as the average MFU across the clusters. 
As shown in~\fig{fig:eval-mfu}, \sys{} consistently surpasses by all baselines under all configurations.
Specifically, all implementations perform similarly with $16$ GPUs since there is less opportunity in small scale. 
With increased CP degrees, Ring Attention falls short due to the decreased compute efficiency as discussed in~\refsec{sec:eval:hopper-attention}; while ByteScale performs even worse due to the significant workload imbalance as shown in~\refsec{sec:eval:balance} under the long-tailed distribution.


\subsection{Ablation Studies}
\label{sec:eval:ablation}
To validate the effectiveness of each proposed design in \sys{}, we conduct ablation studies by adding on components one-by-one, including block-level pipelining (\#1), congestion-free solver (\#2), bottom-up coalescer (\#3), and transparent reshuffler (\#4). 
We use the same evaluation setup as described in~\refsec{sec:eval:setup}, with the number of GPUs fixed at $128\times$ GPU-X for simplicity. 
As shown in~Table~\ref{tab:eval-ablation}, each component contributes to considerable utilization improvement.

\subsection{Sensitivity Tests}
\label{sec:eval:sensitivity-test}
\begin{figure}[t]
    \centering
    \includegraphics[width=\columnwidth]{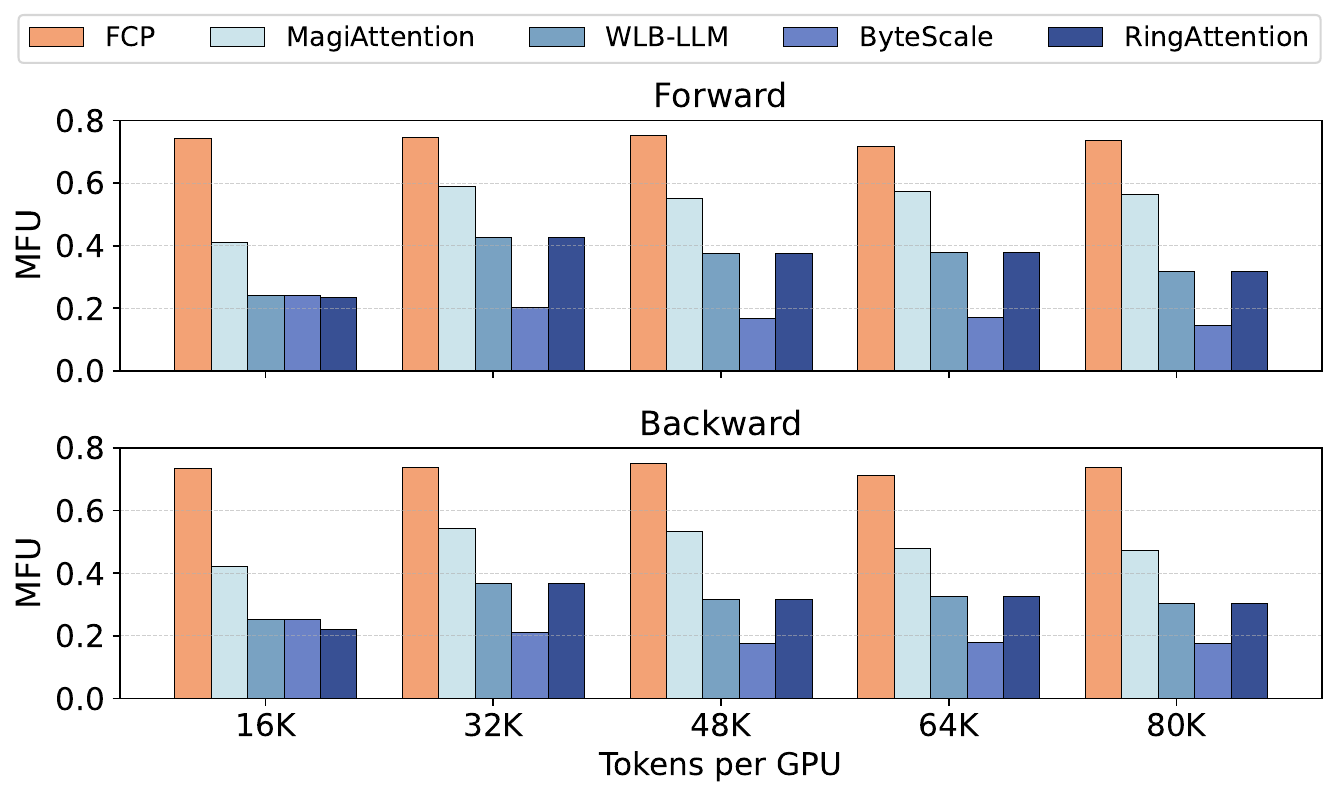}
    \vspace{-4mm}
    \caption{Sensitivity test of per-GPU tokens on $128\times$ GPU-X.}
    \label{fig:eval-sensitivity-tokens}
\end{figure}
\begin{figure}[t]
    \centering
    \includegraphics[width=\columnwidth]{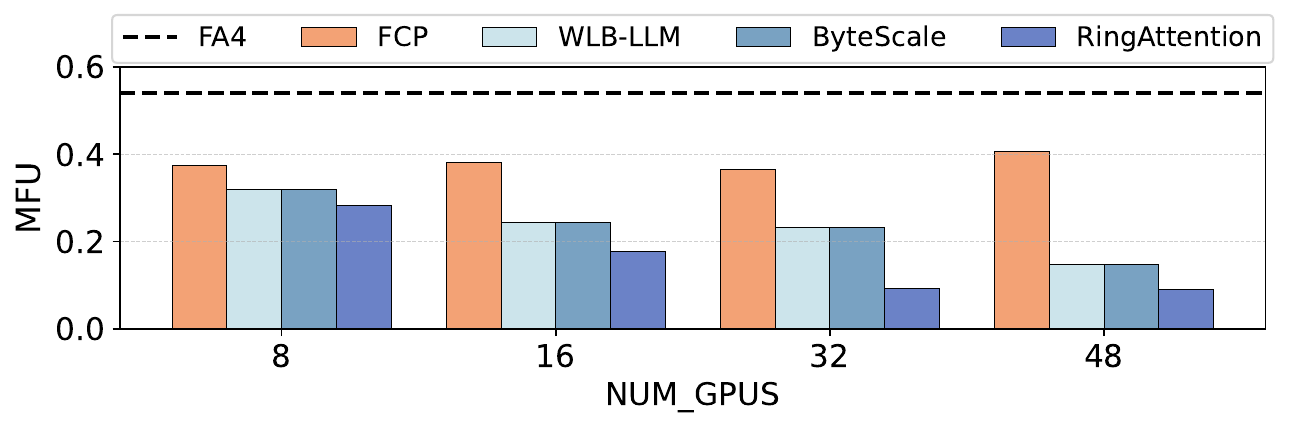}
    \vspace{-4mm}
    \caption{\textit{Weak scaling} test of module-level attention MFU (forward-only), with various number of GPU-Y.}
    \label{fig:eval-sensitivity-gpu}
\end{figure}

\MyPara{Various block size.}
We measure the module-level attention MFU under various block sizes, with the same workload under $128\times$ GPU-X.
As shown in~\fig{fig:eval-sensitivity-block-size}, \sys{} achieves the best MFU with $4$K block size, achieving a sweet spot between workload balance and compute efficiency.

\MyPara{Various number of per-GPU tokens.}
To demonstrate the generality of \sys{}, we also measure the module-level MFU with different number of per-GPU tokens . 
As shown in~\fig{fig:eval-sensitivity-tokens}, \sys{} consistently surpasses all baselines.

\MyPara{GPU types.} 
To assess the portability and generality of \sys{}, we evaluate it on GPU-Y. We adopt the state-of-the-art FlashAttention-4 implementation and adjust the baselines accordingly. We exclude MagiAttention because it relies on customized kernels. As shown in~\fig{fig:eval-sensitivity-gpu}, \sys{} consistently outperforms the baselines, achieving over $70$\% of the single-GPU FlashAttention-4 MFU. The remaining gap is mainly due to the SM resources consumed by communication and the higher arithmetic intensity required.

\section{Discussion and Limitation}
\label{sec:discussion}

\MyPara{Hyperparameter selection.}
The block size in \sys{} is determined automatically by the cluster configuration and workload characteristics.
Concretely, the block size is chosen to (1) saturate the attention kernel-level MFU as shown in~\fig{fig:compute-efficiency}, (2) enable computation--communication overlap, and (3) balance compute and communication across GPUs.
Given a fixed cluster configuration, the first two factors can be profiled in advance, which yields a hardware-efficient block size (e.g., $2$K on GPU-X).
In practice, these hardware-driven choices provide sufficient scheduling flexibility for production workloads (e.g., $32$K context length in Gemini training~\cite{gemini2024}), offering a simple and effective solution.
Further fine-tuning of the block size for a particular workload provides only marginal improvement.
\fig{fig:eval-sensitivity-block-size} shows about a $7$\% difference between $2$K and $6$K block sizes.
For scenarios where users require the absolute optimal configuration, a cost model-based simulation can efficiently validate load balance by running the LPT-based scheduling without executing the full workload.

\MyPara{Limitations and network assumptions.}
\sys{} requires arbitrary peer-to-peer communication.
When multiple on-the-fly packets are coalesced, the resulting pattern effectively resembles all-to-all communication, which demands a topology well suited for such traffic.
Consequently, \sys{} generalizes well on fat-tree or rail-optimized networks with InfiniBand or RoCE on NVIDIA GPUs, but its performance is limited on topologies such as torus-based TPU v3 clusters.

\section{Related Works}
\label{sec:related-work}
\MyPara{Long-context pre-training.} 
LoongTrain~\cite{gu2024loongtrainefficienttraininglongsequence} proposes a double-ring topology to leverage high intra-node interconnection bandwidth. 
However, its parallelism design is input-agnostic, leading to compute inefficiency on short sequences. 
FlexSP~\cite{wang2025flexspacceleratinglargelanguage} dynamically partitions sequences into different SP groups.
However, it parallelizes along the head dimension, whose scalability is limited. 
Another concurrent work, CAD~\cite{zhuang2025efficientlongcontextlanguagemodel}, introduces a fine-grained, load-balanced CP strategy with attention disaggregation. 
However, it fails to model communication traffic after shard migration, resulting in suboptimal performance. 
In contrast, \sys{} is a fine-grained, input-adaptive CP method that jointly achieves computation and communication balance, ensuring efficient execution through congestion-free communication.

\MyPara{Dynamic attention masks.} 
Another line of concurrent work focuses on load balance in parallelizing block-sparse attention~\cite{xi2025sparsevideogenacceleratingvideo}, including BurstEngine~\cite{sun2025burstengineefficientdistributedframework}, MagiAttention~\cite{magiattention2025}, and DCP~\cite{Jiang_2025}. 
These methods adopt a similar block-level abstraction to allow flexible scheduling across workers. 
However, they fail to consider communication congestion, leading to suboptimal performance. 
Although evaluated on causal masks, the block-level abstraction of \sys{} can be extended to block-sparse by deriving block dependencies from the sparsity map.

\section{Conclusion}
We propose \sys{}, a scalable context parallelism paradigm that achieves both compute efficiency and workload balance. The key idea of \sys{} is to enable flexible sequence partition by leveraging arbitrary peer-to-peer communication. With a communication planner, \sys{} ensures efficient runtime execution. Our evaluation shows that \sys{} scales near-linearly up to 256 GPUs and improves attention MFU by up to $2\times$.

\section*{Acknowledgements}
We thank Jiaming Tang and Yifan Qiao for their insightful discussions and feedback. We also thank Heng Zhang, Yanghua Peng, and Haibin Lin for their support throughout the project.

\nocite{*}

\bibliography{main}

@misc{magi-paper,
      title={MAGI-1: Autoregressive Video Generation at Scale}, 
      author={Hansi Teng and Hongyu Jia and Lei Sun and Lingzhi Li and Maolin Li and Mingqiu Tang and Shuai Han and Tianning Zhang and W. Q. Zhang and Weifeng Luo and Xiaoyang Kang and Yuchen Sun and Yue Cao and Yunpeng Huang and Yutong Lin and Yuxin Fang and Zewei Tao and Zheng Zhang and Zhongshu Wang and Zixun Liu and Dai Shi and Guoli Su and Hanwen Sun and Hong Pan and Jie Wang and Jiexin Sheng and Min Cui and Min Hu and Ming Yan and Shucheng Yin and Siran Zhang and Tingting Liu and Xianping Yin and Xiaoyu Yang and Xin Song and Xuan Hu and Yankai Zhang and Yuqiao Li},
      year={2025},
      eprint={2505.13211},
      archivePrefix={arXiv},
      primaryClass={cs.CV},
      url={https://arxiv.org/abs/2505.13211}, 
}

@misc{zhao2023pytorchfsdpexperiencesscaling,
      title={PyTorch FSDP: Experiences on Scaling Fully Sharded Data Parallel}, 
      author={Yanli Zhao and Andrew Gu and Rohan Varma and Liang Luo and Chien-Chin Huang and Min Xu and Less Wright and Hamid Shojanazeri and Myle Ott and Sam Shleifer and Alban Desmaison and Can Balioglu and Pritam Damania and Bernard Nguyen and Geeta Chauhan and Yuchen Hao and Ajit Mathews and Shen Li},
      year={2023},
      eprint={2304.11277},
      archivePrefix={arXiv},
      primaryClass={cs.DC},
      url={https://arxiv.org/abs/2304.11277}, 
}

@misc{deepseekai2024deepseekv2strongeconomicalefficient,
      title={DeepSeek-V2: A Strong, Economical, and Efficient Mixture-of-Experts Language Model}, 
      author={DeepSeek-AI},
      year={2024},
      eprint={2405.04434},
      archivePrefix={arXiv},
      primaryClass={cs.CL},
      url={https://arxiv.org/abs/2405.04434}, 
}

@article{Csardi_The_igraph_software_2006,
author = {Csárdi, Gábor and Nepusz, Tamás},
journal = {InterJournal, Complex Systems},
pages = {1695},
title = {{The igraph software package for complex network research}},
year = {2006}
}

@misc{shah2024flashattention3fastaccurateattention,
      title={FlashAttention-3: Fast and Accurate Attention with Asynchrony and Low-precision}, 
      author={Jay Shah and Ganesh Bikshandi and Ying Zhang and Vijay Thakkar and Pradeep Ramani and Tri Dao},
      year={2024},
      eprint={2407.08608},
      archivePrefix={arXiv},
      primaryClass={cs.LG},
      url={https://arxiv.org/abs/2407.08608}, 
}

@misc{esser2024scalingrectifiedflowtransformers,
      title={Scaling Rectified Flow Transformers for High-Resolution Image Synthesis}, 
      author={Patrick Esser and Sumith Kulal and Andreas Blattmann and Rahim Entezari and Jonas Müller and Harry Saini and Yam Levi and Dominik Lorenz and Axel Sauer and Frederic Boesel and Dustin Podell and Tim Dockhorn and Zion English and Kyle Lacey and Alex Goodwin and Yannik Marek and Robin Rombach},
      year={2024},
      eprint={2403.03206},
      archivePrefix={arXiv},
      primaryClass={cs.CV},
      url={https://arxiv.org/abs/2403.03206}, 
}

@misc{Hopcroft,
   author = "Wikipedia",
   title = "{Hopcroft–Karp algorithm} --- {W}ikipedia{,} The Free Encyclopedia",
   year = "2025",
   howpublished = {\url{http://en.wikipedia.org/w/index.php?title=Hopcroft\%E2\%80\%93Karp\%20algorithm&oldid=1290392689}},
   note = "[Online; accessed 24-October-2025]"
 }

@article{Brandon2023StripedAF,
  title={Striped Attention: Faster Ring Attention for Causal Transformers},
  author={William Brandon and Aniruddha Nrusimha and Kevin Qian and Zack Ankner and Tian Jin and Zhiye Song and Jonathan Ragan-Kelley},
  journal={ArXiv},
  year={2023},
  volume={abs/2311.09431},
  url={https://api.semanticscholar.org/CorpusID:265220849}
}

@misc{vaswani2023attentionneed,
      title={Attention Is All You Need}, 
      author={Ashish Vaswani and Noam Shazeer and Niki Parmar and Jakob Uszkoreit and Llion Jones and Aidan N. Gomez and Lukasz Kaiser and Illia Polosukhin},
      year={2023},
      eprint={1706.03762},
      archivePrefix={arXiv},
      primaryClass={cs.CL},
      url={https://arxiv.org/abs/1706.03762}, 
}

@misc{osama2023streamkworkcentricparalleldecomposition,
      title={Stream-K: Work-centric Parallel Decomposition for Dense Matrix-Matrix Multiplication on the GPU}, 
      author={Muhammad Osama and Duane Merrill and Cris Cecka and Michael Garland and John D. Owens},
      year={2023},
      eprint={2301.03598},
      archivePrefix={arXiv},
      primaryClass={cs.DS},
      url={https://arxiv.org/abs/2301.03598}, 
}

@inproceedings{dao2023flashattention2,
  title={Flash{A}ttention-2: Faster Attention with Better Parallelism and Work Partitioning},
  author={Dao, Tri},
  booktitle={International Conference on Learning Representations (ICLR)},
  year={2024}
}

@misc{cui2025optimizingsloorientedllmserving,
      title={Optimizing SLO-oriented LLM Serving with PD-Multiplexing}, 
      author={Weihao Cui and Yukang Chen and Han Zhao and Ziyi Xu and Quan Chen and Xusheng Chen and Yangjie Zhou and Shixuan Sun and Minyi Guo},
      year={2025},
      eprint={2504.14489},
      archivePrefix={arXiv},
      primaryClass={cs.OS},
      url={https://arxiv.org/abs/2504.14489}, 
}

@article{ye2025flashinfer,
    title = {FlashInfer: Efficient and Customizable Attention Engine for LLM Inference Serving},
    author = {
      Ye, Zihao and
      Chen, Lequn and
      Lai, Ruihang and
      Lin, Wuwei and
      Zhang, Yineng and
      Wang, Stephanie and
      Chen, Tianqi and
      Kasikci, Baris and
      Grover, Vinod and
      Krishnamurthy, Arvind and
      Ceze, Luis
    },
    journal = {arXiv preprint arXiv:2501.01005},
    year = {2025},
    url = {https://arxiv.org/abs/2501.01005}
}

@inproceedings{MLSYS2024_5edb57c0,
 author = {Zhao, Yilong and Lin, Chien-Yu and Zhu, Kan and Ye, Zihao and Chen, Lequn and Zheng, Size and Ceze, Luis and Krishnamurthy, Arvind and Chen, Tianqi and Kasikci, Baris},
 booktitle = {Proceedings of Machine Learning and Systems},
 editor = {P. Gibbons and G. Pekhimenko and C. De Sa},
 pages = {196--209},
 title = {Atom: Low-Bit Quantization for Efficient and Accurate LLM Serving},
 url = {https://proceedings.mlsys.org/paper_files/paper/2024/file/5edb57c05c81d04beb716ef1d542fe9e-Paper-Conference.pdf},
 volume = {6},
 year = {2024}
}

@inproceedings{dao2022flashattention,
  title={Flash{A}ttention: Fast and Memory-Efficient Exact Attention with {IO}-Awareness},
  author={Dao, Tri and Fu, Daniel Y. and Ermon, Stefano and Rudra, Atri and R{\'e}, Christopher},
  booktitle={Advances in Neural Information Processing Systems (NeurIPS)},
  year={2022}
}

@misc{fang2024uspunifiedsequenceparallelism,
      title={USP: A Unified Sequence Parallelism Approach for Long Context Generative AI}, 
      author={Jiarui Fang and Shangchun Zhao},
      year={2024},
      eprint={2405.07719},
      archivePrefix={arXiv},
      primaryClass={cs.LG},
      url={https://arxiv.org/abs/2405.07719}, 
}

@misc{jacobs2023deepspeedulyssesoptimizationsenabling,
      title={DeepSpeed Ulysses: System Optimizations for Enabling Training of Extreme Long Sequence Transformer Models}, 
      author={Sam Ade Jacobs and Masahiro Tanaka and Chengming Zhang and Minjia Zhang and Shuaiwen Leon Song and Samyam Rajbhandari and Yuxiong He},
      year={2023},
      eprint={2309.14509},
      archivePrefix={arXiv},
      primaryClass={cs.LG},
      url={https://arxiv.org/abs/2309.14509}, 
}

@misc{wang2024railonlylowcosthighperformancenetwork,
      title={Rail-only: A Low-Cost High-Performance Network for Training LLMs with Trillion Parameters}, 
      author={Weiyang Wang and Manya Ghobadi and Kayvon Shakeri and Ying Zhang and Naader Hasani},
      year={2024},
      eprint={2307.12169},
      archivePrefix={arXiv},
      primaryClass={cs.NI},
      url={https://arxiv.org/abs/2307.12169}, 
}

@misc{LPT,
   author = "Wikipedia",
   title = "{Longest-processing-time-first scheduling} --- {W}ikipedia{,} The Free Encyclopedia",
   year = "2025",
   howpublished = {\url{http://en.wikipedia.org/w/index.php?title=Longest-processing-time-first\%20scheduling&oldid=1315996044}},
   note = "[Online; accessed 21-October-2025]"
 }

@misc{Sivamani2025NVIDIA,
	author = {Sivamani, Kirthi Shankar and Moon, Tim and Tredak, Przemyslaw and Yang, Charlene and Nguyen, Phuong},
	year = {2025},
	month = {oct 28},
	title = {NVIDIA/{TransformerEngine}},
	url = {https://github.com/NVIDIA/TransformerEngine},
	howpublished = {https://github.com/NVIDIA/TransformerEngine},
}

@misc{cameron2025hallsmarriagetheorem,
      title={Hall's marriage theorem}, 
      author={Peter J. Cameron},
      year={2025},
      eprint={2503.23159},
      archivePrefix={arXiv},
      primaryClass={math.CO},
      url={https://arxiv.org/abs/2503.23159}, 
}

@misc{h20,
	author = {Dylan Patel},
	title = {{N}vidia's {N}ew {C}hina {A}{I} {C}hips {C}ircumvent {U}{S} {R}estrictions | {H}20 {F}aster {T}han {H}100 | {H}uawei {A}scend 910{B} --- newsletter.semianalysis.com},
	howpublished = {\url{https://newsletter.semianalysis.com/p/nvidias-new-china-ai-chips-circumvent}},
	year = {2025},
	note = {[Accessed 28-10-2025]},
}

@misc{widenBlackwelldatasheet3384703pdf,
	author = {NVIDIA},
	title = {NVL72-Datasheet},
	howpublished = {\url{https://nvdam.widen.net/s/wwnsxrhm2w/blackwell-datasheet-3384703}},
	year = {2025},
	note = {[Accessed 28-10-2025]},
}

@misc{yang2025qwen3technicalreport,
      title={Qwen3 Technical Report}, 
      author={An Yang and Anfeng Li and Baosong Yang and Beichen Zhang and Binyuan Hui and Bo Zheng and Bowen Yu and Chang Gao and Chengen Huang and Chenxu Lv and Chujie Zheng and Dayiheng Liu and Fan Zhou and Fei Huang and Feng Hu and Hao Ge and Haoran Wei and Huan Lin and Jialong Tang and Jian Yang and Jianhong Tu and Jianwei Zhang and Jianxin Yang and Jiaxi Yang and Jing Zhou and Jingren Zhou and Junyang Lin and Kai Dang and Keqin Bao and Kexin Yang and Le Yu and Lianghao Deng and Mei Li and Mingfeng Xue and Mingze Li and Pei Zhang and Peng Wang and Qin Zhu and Rui Men and Ruize Gao and Shixuan Liu and Shuang Luo and Tianhao Li and Tianyi Tang and Wenbiao Yin and Xingzhang Ren and Xinyu Wang and Xinyu Zhang and Xuancheng Ren and Yang Fan and Yang Su and Yichang Zhang and Yinger Zhang and Yu Wan and Yuqiong Liu and Zekun Wang and Zeyu Cui and Zhenru Zhang and Zhipeng Zhou and Zihan Qiu},
      year={2025},
      eprint={2505.09388},
      archivePrefix={arXiv},
      primaryClass={cs.CL},
      url={https://arxiv.org/abs/2505.09388}, 
}

@misc{grattafiori2024llama3herdmodels,
      title={The Llama 3 Herd of Models}, 
      author={Meta AI},
      year={2024},
      eprint={2407.21783},
      archivePrefix={arXiv},
      primaryClass={cs.AI},
      url={https://arxiv.org/abs/2407.21783}, 
}

@misc{chandran2024resultslptnearlineartime,
      title={Two Results on LPT: A Near-Linear Time Algorithm and Parcel Delivery using Drones}, 
      author={L. Sunil Chandran and Rishikesh Gajjala and Shravan Mehra and Saladi Rahul},
      year={2024},
      eprint={2407.16323},
      archivePrefix={arXiv},
      primaryClass={cs.DS},
      url={https://arxiv.org/abs/2407.16323}, 
}

@misc{ainslie2023gqatraininggeneralizedmultiquery,
      title={GQA: Training Generalized Multi-Query Transformer Models from Multi-Head Checkpoints}, 
      author={Joshua Ainslie and James Lee-Thorp and Michiel de Jong and Yury Zemlyanskiy and Federico Lebrón and Sumit Sanghai},
      year={2023},
      eprint={2305.13245},
      archivePrefix={arXiv},
      primaryClass={cs.CL},
      url={https://arxiv.org/abs/2305.13245}, 
}

@misc{shoeybi2020megatronlmtrainingmultibillionparameter,
      title={Megatron-LM: Training Multi-Billion Parameter Language Models Using Model Parallelism}, 
      author={Mohammad Shoeybi and Mostofa Patwary and Raul Puri and Patrick LeGresley and Jared Casper and Bryan Catanzaro},
      year={2020},
      eprint={1909.08053},
      archivePrefix={arXiv},
      primaryClass={cs.CL},
      url={https://arxiv.org/abs/1909.08053}, 
}

@misc{magiattention2025,
  title={MagiAttention: A Distributed Attention Towards Linear Scalability for Ultra-Long Context, Heterogeneous Mask Training},
  author={Zewei, Tao and Yunpeng, Huang},
  year={2025},
  howpublished={\url{https://github.com/SandAI-org/MagiAttention/}},
}

@misc{chen2023punicamultitenantloraserving,
      title={Punica: Multi-Tenant LoRA Serving}, 
      author={Lequn Chen and Zihao Ye and Yongji Wu and Danyang Zhuo and Luis Ceze and Arvind Krishnamurthy},
      year={2023},
      eprint={2310.18547},
      archivePrefix={arXiv},
      primaryClass={cs.DC},
      url={https://arxiv.org/abs/2310.18547}, 
}

@misc{yang2025contextparallelismscalablemilliontoken,
      title={Context Parallelism for Scalable Million-Token Inference}, 
      author={Amy Yang and Jingyi Yang and Aya Ibrahim and Xinfeng Xie and Bangsheng Tang and Grigory Sizov and Jeremy Reizenstein and Jongsoo Park and Jianyu Huang},
      year={2025},
      eprint={2411.01783},
      archivePrefix={arXiv},
      primaryClass={cs.DC},
      url={https://arxiv.org/abs/2411.01783}, 
}

@misc{zheng2022alpaautomatinginterintraoperator,
      title={Alpa: Automating Inter- and Intra-Operator Parallelism for Distributed Deep Learning}, 
      author={Lianmin Zheng and Zhuohan Li and Hao Zhang and Yonghao Zhuang and Zhifeng Chen and Yanping Huang and Yida Wang and Yuanzhong Xu and Danyang Zhuo and Eric P. Xing and Joseph E. Gonzalez and Ion Stoica},
      year={2022},
      eprint={2201.12023},
      archivePrefix={arXiv},
      primaryClass={cs.LG},
      url={https://arxiv.org/abs/2201.12023}, 
}

@misc{pureai_cost,
  author       = {David Ramel},
  title        = {Open Source Foundation Models Cost Tens of Millions to Train},
  howpublished = {\url{https://pureai.com/articles/2024/04/23/open-source-models-cost.aspx}},
  year         = {2024},
  note         = {Accessed: 2025-10-07}
}

@article{severson2024trainingcost,
  author  = {Severson, Matthew and others},
  title   = {The Cost of Training Frontier AI Models},
  journal = {arXiv preprint arXiv:2405.21015},
  year    = {2024}
}

@misc{apple_foundation,
  author       = {Apple Machine Learning Research},
  title        = {Apple Foundation Models: 2025 Updates},
  howpublished = {\url{https://machinelearning.apple.com/research/apple-foundation-models-2025-updates}},
  year         = {2025},
  note         = {Accessed: 2025-10-07}
}

@misc{meta_longcontext,
  author       = {Meta AI Research},
  title        = {Effective Long-Context Scaling of Foundation Models},
  howpublished = {\url{https://ai.meta.com/research/publications/effective-long-context-scaling-of-foundation-models/}},
  year         = {2024},
  note         = {Accessed: 2025-10-07}
}

@inproceedings{Jiang_2025, series={SOSP ’25},
   title={DCP: Addressing Input Dynamism In Long-Context Training via Dynamic Context Parallelism},
   url={http://dx.doi.org/10.1145/3731569.3764849},
   DOI={10.1145/3731569.3764849},
   booktitle={Proceedings of the ACM SIGOPS 31st Symposium on Operating Systems Principles},
   publisher={ACM},
   author={Jiang, Chenyu and Cai, Zhenkun and Tian, Ye and Jia, Zhen and Wang, Yida and Wu, Chuan},
   year={2025},
   month=oct, pages={221–236},
   collection={SOSP ’25} }

@misc{sun2025burstengineefficientdistributedframework,
      title={BurstEngine: an Efficient Distributed Framework for Training Transformers on Extremely Long Sequences of over 1M Tokens}, 
      author={Ao Sun and Weilin Zhao and Xu Han and Cheng Yang and Zhiyuan Liu and Chuan Shi and Maosong sun},
      year={2025},
      eprint={2509.19836},
      archivePrefix={arXiv},
      primaryClass={cs.DC},
      url={https://arxiv.org/abs/2509.19836}, 
}

@misc{xi2025sparsevideogenacceleratingvideo,
      title={Sparse VideoGen: Accelerating Video Diffusion Transformers with Spatial-Temporal Sparsity}, 
      author={Haocheng Xi and Shuo Yang and Yilong Zhao and Chenfeng Xu and Muyang Li and Xiuyu Li and Yujun Lin and Han Cai and Jintao Zhang and Dacheng Li and Jianfei Chen and Ion Stoica and Kurt Keutzer and Song Han},
      year={2025},
      eprint={2502.01776},
      archivePrefix={arXiv},
      primaryClass={cs.CV},
      url={https://arxiv.org/abs/2502.01776}, 
}

@misc{zhuang2025efficientlongcontextlanguagemodel,
      title={Efficient Long-context Language Model Training by Core Attention Disaggregation}, 
      author={Yonghao Zhuang and Junda Chen and Bo Pang and Yi Gu and Yibo Zhu and Yimin Jiang and Ion Stoica and Eric Xing and Hao Zhang},
      year={2025},
      eprint={2510.18121},
      archivePrefix={arXiv},
      primaryClass={cs.LG},
      url={https://arxiv.org/abs/2510.18121}, 
}

@misc{magi-bench,
	author = {Zewei Tao and Yunpeng Huang	},
	title = {Official Benchmark Scripts for MagiAttention},
	howpublished = {\url{https://github.com/SandAI-org/MagiAttention/blob/main/exps/dist_attn/run_benchmark.py}},
	year = {2025},
	note = {[Accessed 28-10-2025]},
}

@misc{githubGitHubAshZhengWLBLLMCP,
	author = {Zheng Wang},
	title = {Official WLB-LLM Codebase},
	howpublished = {\url{https://github.com/Ash-Zheng/WLB-LLM-CP}},
	year = {2025},
	note = {[Accessed 28-10-2025]},
}

@article{ringattention_arxiv,
  author  = {H. Liu and others},
  title   = {Ring Attention with Blockwise Transformers for Near-Infinite Context},
  journal = {arXiv preprint arXiv:2310.01889},
  year    = {2023},
  url     = {https://arxiv.org/abs/2310.01889}
}

@article{goyal2017imagenet,
  author  = {Goyal, Priya and Doll{\'a}r, Piotr and Girshick, Ross and Noordhuis, Pieter and Wesolowski, Lukasz and Kyrola, Aapo and Tulloch, Andrew and Jia, Yangqing and He, Kaiming},
  title   = {Accurate, Large Minibatch {SGD}: Training ImageNet in 1 Hour},
  journal = {arXiv preprint arXiv:1706.02677},
  year    = {2017},
  url     = {https://arxiv.org/abs/1706.02677}
}

@article{sergeev2018horovod,
  author  = {Sergeev, Alexander and Del Balso, Mike},
  title   = {Horovod: Fast and Easy Distributed Deep Learning in TensorFlow},
  journal = {arXiv preprint arXiv:1802.05799},
  year    = {2018},
  url     = {https://arxiv.org/abs/1802.05799}
}

@article{shoeybi2019megatron,
  author  = {Shoeybi, Mohammad and Patwary, Mostofa and Puri, Raul and LeGresley, Patrick and Casper, Jared and Catanzaro, Bryan},
  title   = {Megatron-{LM}: Training Multi-Billion Parameter Language Models Using Model Parallelism},
  journal = {arXiv preprint arXiv:1909.08053},
  year    = {2019},
  url     = {https://arxiv.org/abs/1909.08053}
}

@inproceedings{Ge_2025, series={SIGCOMM ’25},
   title={ByteScale: Communication-Efficient Scaling of LLM Training with a 2048K Context Length on 16384 GPUs},
   url={http://dx.doi.org/10.1145/3718958.3754352},
   DOI={10.1145/3718958.3754352},
   booktitle={Proceedings of the ACM SIGCOMM 2025 Conference},
   publisher={ACM},
   author={Ge, Hao and Feng, Junda and Huang, Qi and Fu, Fangcheng and Nie, Xiaonan and Zuo, Lei and Lin, Haibin and Cui, Bin and Liu, Xin},
   year={2025},
   month=aug, pages={963–978},
   collection={SIGCOMM ’25} }

@misc{gu2024loongtrainefficienttraininglongsequence,
      title={LoongTrain: Efficient Training of Long-Sequence LLMs with Head-Context Parallelism}, 
      author={Diandian Gu and Peng Sun and Qinghao Hu and Ting Huang and Xun Chen and Yingtong Xiong and Guoteng Wang and Qiaoling Chen and Shangchun Zhao and Jiarui Fang and Yonggang Wen and Tianwei Zhang and Xin Jin and Xuanzhe Liu},
      year={2024},
      eprint={2406.18485},
      archivePrefix={arXiv},
      primaryClass={cs.DC},
      url={https://arxiv.org/abs/2406.18485}, 
}

@inproceedings{wlb-llm,
author = {Wang, Zheng and Cai, Anna and Xie, Xinfeng and Pan, Zaifeng and Guan, Yue and Chu, Weiwei and Wang, Jie and Li, Shikai and Huang, Jianyu and Cai, Chris and Hao, Yuchen and Ding, Yufei},
title = {WLB-LLM: workload-balanced 4D parallelism for large language model training},
year = {2025},
isbn = {978-1-939133-47-2},
publisher = {USENIX Association},
address = {USA},
booktitle = {Proceedings of the 19th USENIX Conference on Operating Systems Design and Implementation},
articleno = {43},
numpages = {17},
location = {Boston, MA, USA},
series = {OSDI '25}
}

@misc{wang2025flexspacceleratinglargelanguage,
      title={FlexSP: Accelerating Large Language Model Training via Flexible Sequence Parallelism}, 
      author={Yujie Wang and Shiju Wang and Shenhan Zhu and Fangcheng Fu and Xinyi Liu and Xuefeng Xiao and Huixia Li and Jiashi Li and Faming Wu and Bin Cui},
      year={2025},
      eprint={2412.01523},
      archivePrefix={arXiv},
      primaryClass={cs.DC},
      url={https://arxiv.org/abs/2412.01523}, 
}

@misc{li2024distflashattndistributedmemoryefficientattention,
      title={DISTFLASHATTN: Distributed Memory-efficient Attention for Long-context LLMs Training}, 
      author={Dacheng Li and Rulin Shao and Anze Xie and Eric P. Xing and Xuezhe Ma and Ion Stoica and Joseph E. Gonzalez and Hao Zhang},
      year={2024},
      eprint={2310.03294},
      archivePrefix={arXiv},
      primaryClass={cs.LG},
      url={https://arxiv.org/abs/2310.03294}, 
}

@article{ravikumar2024straggler,
  author  = {Ravikumar, A. and Parthasarathy, S. and Thyagarajan, K. and others},
  title   = {DPro\textendash SM: A Distributed Framework for Proactive Straggler Mitigation in Synchronous {SGD}},
  journal = {Heliyon},
  year    = {2024},
  note    = {Stragglers delay synchronization during each iteration's aggregation step},
  url     = {https://www.sciencedirect.com/science/article/pii/S2405844023107754}
}

@inproceedings{narayanan2021pipedreamflush,
  author    = {Narayanan, Deepak and Shoeybi, Mohammad and Cho, Tushar and others},
  title     = {Memory-Efficient Pipeline-Parallel {DNN} Training},
  booktitle = {Proceedings of the 38th International Conference on Machine Learning (ICML)},
  year      = {2021},
  url       = {https://proceedings.mlr.press/v139/narayanan21a/narayanan21a.pdf}
}

@article{rajbhandari2020zero,
  author  = {Rajbhandari, Samyam and Rasley, Jeff and Ruwase, Olatunji and He, Yuxiong},
  title   = {{ZeRO}: Memory Optimizations Toward Training Trillion Parameter Models},
  journal = {arXiv preprint arXiv:1910.02054},
  year    = {2020},
  url     = {https://arxiv.org/abs/1910.02054}
}

@misc{gao2025seedance10exploringboundaries,
      title={Seedance 1.0: Exploring the Boundaries of Video Generation Models}, 
      author={Yu Gao and Haoyuan Guo and Tuyen Hoang and Weilin Huang and Lu Jiang and Fangyuan Kong and Huixia Li and Jiashi Li and Liang Li and Xiaojie Li and Xunsong Li and Yifu Li and Shanchuan Lin and Zhijie Lin and Jiawei Liu and Shu Liu and Xiaonan Nie and Zhiwu Qing and Yuxi Ren and Li Sun and Zhi Tian and Rui Wang and Sen Wang and Guoqiang Wei and Guohong Wu and Jie Wu and Ruiqi Xia and Fei Xiao and Xuefeng Xiao and Jiangqiao Yan and Ceyuan Yang and Jianchao Yang and Runkai Yang and Tao Yang and Yihang Yang and Zilyu Ye and Xuejiao Zeng and Yan Zeng and Heng Zhang and Yang Zhao and Xiaozheng Zheng and Peihao Zhu and Jiaxin Zou and Feilong Zuo},
      year={2025},
      eprint={2506.09113},
      archivePrefix={arXiv},
      primaryClass={cs.CV},
      url={https://arxiv.org/abs/2506.09113}, 
}

@misc{gemini2024,
  title={Gemini Technical Report},
  author={Gemini Team},
  year={2024},
  eprint={2312.11805},
  archivePrefix={arXiv},
  url={https://arxiv.org/pdf/2312.11805}
}

@misc{cuda_green_ctx,
  title={{CUDA} Driver {API}: Green Contexts},
  author={{NVIDIA}},
  year={2024},
  howpublished={\url{https://docs.nvidia.com/cuda/cuda-driver-api/group__CUDA__GREEN__CONTEXTS.html}}
}

@article{nie2023flexmoe,
  title={Flexmoe: Scaling large-scale sparse pre-trained model training via dynamic device placement},
  author={Nie, Xiaonan and Miao, Xupeng and Wang, Zilong and Yang, Zichao and Xue, Jilong and Ma, Lingxiao and Cao, Gang and Cui, Bin},
  journal={Proceedings of the ACM on Management of Data},
  volume={1},
  number={1},
  pages={1--19},
  year={2023},
  publisher={ACM New York, NY, USA}
}
\bibliographystyle{mlsys2025}

\afterpage{\clearpage}
\newpage
\appendix
\section{Appendix}
\label{sec:appendix}

\subsection{Block Assignment Algorithm}
\label{sec-appendix-assignment-policy}
Given a set of blocks $B$ from different sequences, each block $i$ may have a different computation cost $c_i$. 
This difference arises because, although Zig-Zag packing balances workloads among blocks from the same sequence, blocks from different sequences exhibit distinct arithmetic intensity, leading to varying compute usages.

Thus, \sys{} adopts a variant of the Longest Processing Time (LPT) scheduling algorithm~\cite{chandran2024resultslptnearlineartime}, to reduce workload imbalance. 
The problem is modeled as a multi-dimensional bin-packing problem, where the objective is to minimize the maximum computation usage among all workers, subject to a per-worker memory constraint $M$.

As shown in Algorithm~\ref{alg-design-balance}, the blocks are first sorted by their compute and memory usages (line $2$), and then greedily assigned to workers. 
At each iteration, the worker with the smallest current load is selected. The load is defined as a weighted sum of memory and compute usage, with $\alpha$ and $\beta$ controlling the preference. 
Only workers that satisfy the memory constraint are eligible for selection (line $5$).

\begin{algorithm}[h]
\caption{Greedy Load-Balanced Assignment}
\label{alg-design-balance}
\begin{algorithmic}[1]
\INPUT List of blocks $B$ with memory usages $m_i$ and compute amount $c_i$, number of workers $N$, weights $\alpha$, $\beta$, and memory usage limit $M$, tolerant factor $\delta$
\STATE Compute desired (average) memory and compute loads per worker: $\hat{m} \gets \sum_i m_i / N$, $\hat{c} \gets \sum_i c_i / N$
\STATE Sort blocks in descending order by $\max(m_i / \hat{m},\; c_i / \hat{c})$
\STATE Initialize empty workers $w_1, \ldots, w_N$ with zero workload $(m_w, c_w) = (0, 0)$
\FOR{each block $B_i$ in descending order}
    \STATE Find worker $w^*$ with minimal load and below memory limit ($(m_{w*}+m_i)\le(M*(1+\delta))$):
    \[
    w^* \gets \arg\min_w \max\left( \alpha \cdot \frac{m_w + m_i}{\hat{m}},\; \beta \cdot \frac{c_w + c_i}{\hat{c}} \right)
    \]
    \STATE Assign $B_i$ to $w^*$ and update $(m_{w^*}, c_{w^*})$
\ENDFOR
\end{algorithmic}
\end{algorithm}

\subsection{Proof of Congestion-Free Solver}
\label{sec-appendix-solver}
Given an undirected bipartite graph $G$ with a maximum degree $\Delta$, edge set $E$, and $2N$ vertices, the minimum number of disjoint graph matchings from a decomposition of $E$ is denoted as $\Delta^*$.
As shown in Lemma~\ref{lem:num-matching}, the lower bound of $\Delta^*$ is $\Delta$. Thus, by constructing a decomposition of $E$ into $\Delta$ disjoint matchings $M$, we can prove that $\Delta^* = \Delta$.

We first convert $G$ into a $\Delta$-regular bipartite graph $\hat{G}$ by greedily adding edges into $E$. 
Because the sum of degrees on the first $N$ send nodes equals the sum of degrees on the $N$ receive nodes, it is apparent that an edge $\hat{e}$ from the $i$-th send node to the $j$-th receive node can be inserted where both $\texttt{deg}(i) < \Delta$ and $\texttt{deg}(j) < \Delta$.
We keep adding such edges, denoted as $\hat{E}$, until all nodes reach degree $\Delta$. Therefore, the augmented graph $\hat{G}$ with edge set $E \cup \hat{E}$ becomes a $\Delta$-regular bipartite graph.

By Hall’s theorem~\cite{cameron2025hallsmarriagetheorem}, every regular bipartite graph has a perfect matching. Hence, $\hat{G}$ can be decomposed into $\Delta$ perfect matchings $\hat{M}$ by recursively removing a perfect matching from $\hat{G}$.
Finally, based on the perfect matchings $\hat{M}$, the original matching set $M$ can be obtained by deleting all newly added edges $\hat{E}$. Proof completes.

\subsection{Detailed Configurations of Baselines}
\label{sec-appendix-baselines}
\MyPara{Ring Attention.} 
We implement Ring Attention on top of TransformerEngine~\cite{Sivamani2025NVIDIA}, which follows a peer-to-peer communication schema. 
To enable computation and communication overlap, we adopt double buffering, which allows concurrent operations on two CUDA streams. 
Zig-Zag packing is enabled by default to maintain intra-sequence workload balance.

\MyPara{ByteScale.} 
We reproduce the CP algorithm described in Algorithm~2 of their paper~\cite{Ge_2025}, also known as HDP-balanced, by referring to their private codebases. 
Excluding pipeline parallelism, HDP-balanced assigns sequences to workers in proportion to their context lengths. 
For instance, a sequence of length $k \cdot L$ is assigned to $k\times$ more workers than a sequence of length $L$. 
Within each partition, Ring Attention is applied recursively.

\MyPara{WLB-LLM.} 
Since the official implementation~\cite{githubGitHubAshZhengWLBLLMCP} uses FlashAttention2, which performs suboptimally on latest Hopper and Blackwell GPUs, we reimplement WLB-LLM ourselves. 
We use the maximum performance of Ring Attention and ByteScale as an oracle version, where the online estimator is replaced with an oracle one. 
Note that per-sequence and per-document sharding correspond to ByteScale and Ring Attention, respectively.

\MyPara{MagiAttention.} 
We directly adopt the official codebase implementation\footnote{Commit hash: 3a96ab7}~\cite{magiattention2025}. 
For benchmarking, we use the scripts provided by the authors~\cite{magi-bench}. 
To enable communication and computation overlap, we manually set \texttt{CUDA\_DEVICE\_MAX\_CONNECTIONS} to $8$. 
We also mannuly disable \texttt{MAGI\_ATTENTION\_HIERARCHICAL\_COMM} as it causes NCCL errors in our clusters.


\begin{figure*}[!t]
    \centering
    \begin{minipage}{0.48\textwidth}
        \centering
        \includegraphics[width=\textwidth]{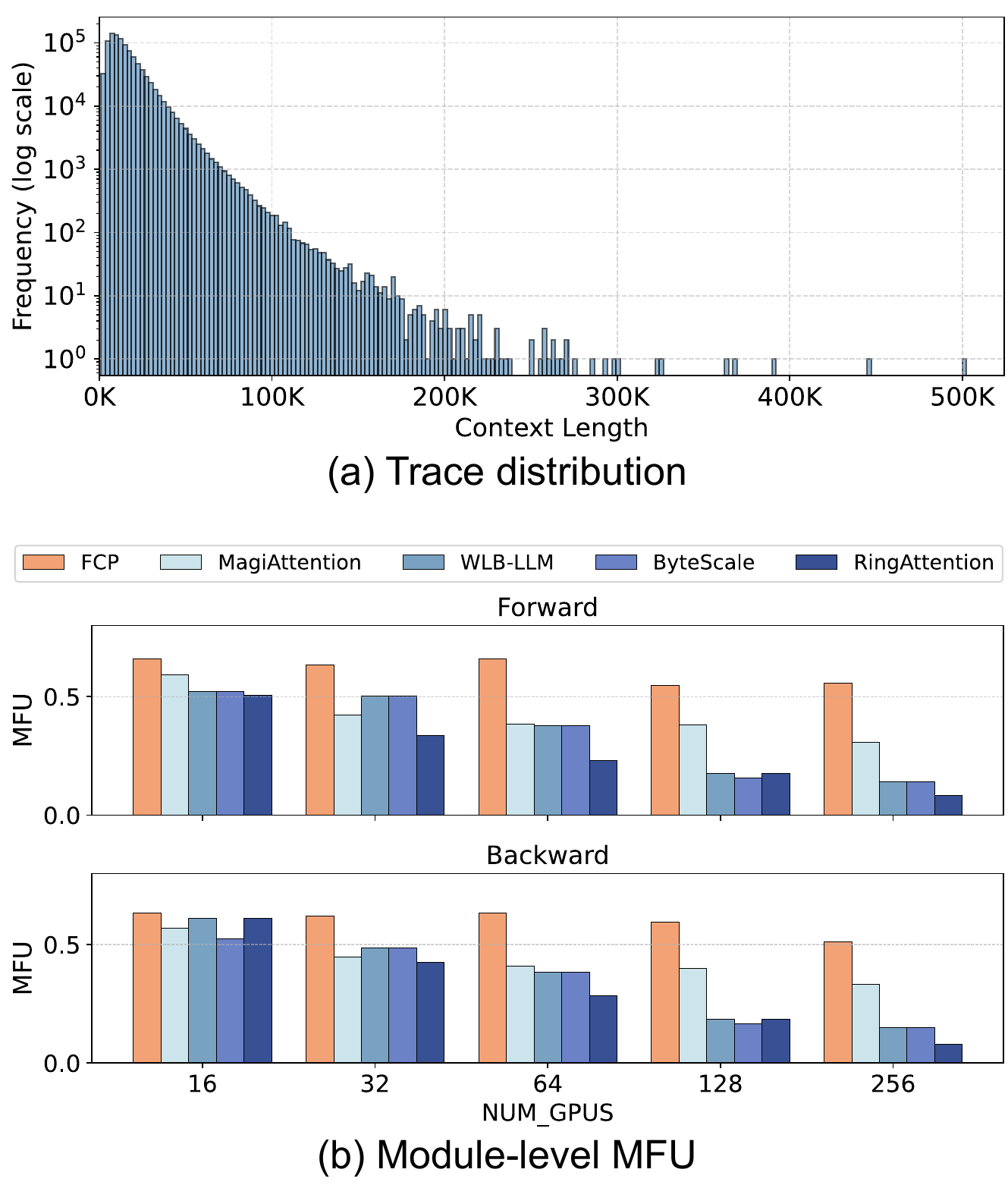}
        \caption{(a) Trace distribution of the \textit{lognormal} distribution and (b) weak-scaling of module-level attention MFU. The number of tokens per GPU is fixed at $32$K.}
        \label{fig:appendix-mfu-lognorm}
    \end{minipage}
    \hfill
    \begin{minipage}{0.48\textwidth}
        \centering
        \includegraphics[width=\textwidth]{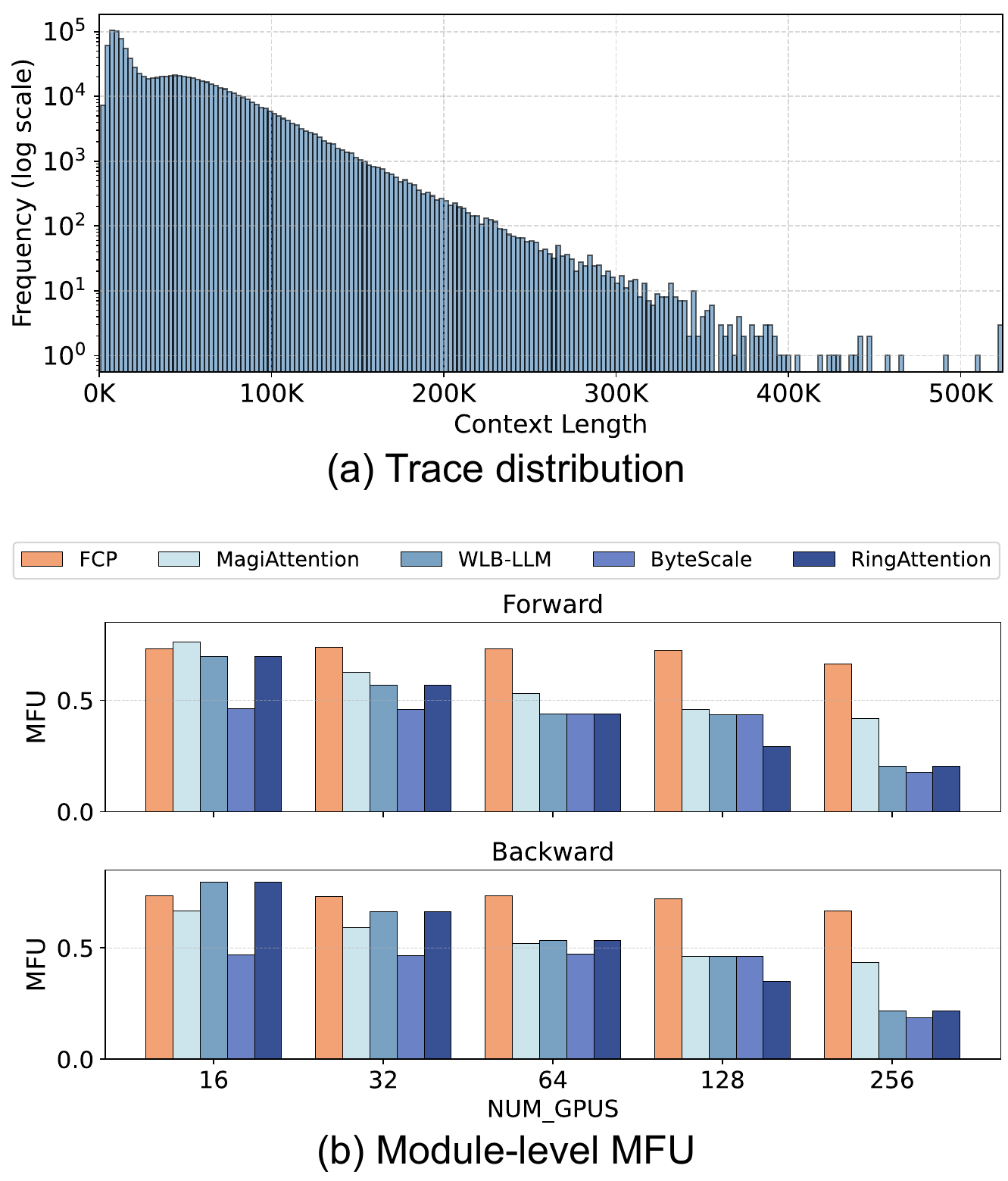}
        \caption{(a) Trace distribution of the \textit{bimodal} distribution and (b) weak-scaling of module-level attention MFU. The number of tokens per GPU is fixed at $32$K.}
        \label{fig:appendix-mfu-bimodal}
    \end{minipage}
\end{figure*}

\subsection{Evaluation on Additional Workloads}
\label{sec-appendix-more-workload}
Besides the real-world distribution derived from our pretraining tasks~\fig{fig:input-distribution}, we also construct two synthetic workloads to demonstrate the generality of \sys{}. 
One exhibits a less long-tailed distribution, and the other follows a bimodal pattern. 
Both are generated by sampling from lognormal distributions. 
The minimum and maximum sequence lengths are set between $1$K and $512$K. 
The remaining evaluation setup is identical to that in~\refsec{sec:eval:setup}.

\MyPara{Less long-tailed.} 
We generate this workload using a lognormal distribution with a standard deviation of $s=0.7$ and an expected sequence length of $16$K. 
As shown in~\fig{fig:appendix-mfu-lognorm} (a), the resulting trace is much more concentrated compared to~\fig{fig:input-distribution}. 
As the extremely long context is rare in these cases, ByteScale performs better than Ring Attention due to less workload imbalance. 
All remaining methods perform worse as the total computation amount is smaller.

\MyPara{Bimodal.} 
We synthesize this workload by combining two lognormal distributions with different mean sequence lengths to form a bimodal pattern. 
The two components are configured as $s=0.5$ with an average of $16$K, and $s=0.5$ with an average of $64$K, as illustrated in~\fig{fig:appendix-mfu-bimodal} (a).


\end{document}